\documentclass[useAMS,usenatbib]{mn2e}
\usepackage{multirow}
\usepackage{threeparttable}

\usepackage{epsfig}
\usepackage{epstopdf}
\usepackage{graphicx,amssymb,supertabular,longtable,lscape} %rotating,
\usepackage{amsmath}
\usepackage{textcomp}
\usepackage{colortbl, xcolor}
\usepackage{multirow}
\usepackage{threeparttable}
\usepackage{color}
\usepackage{tabularx}

\definecolor{mycol}{cmyk}{0.0, 0.11, 0.12, 0.0}
\definecolor{mycol}{cmyk}{0.1, 0.1, 0.1, 0.0}

\def\kms{km~s$^{-1}$}

\def\msun{\mbox{M$_{\odot}$}}
\def\gray{\rowcolor[gray]{0.9}}

\newcommand{\Epic}{{\sc Epic5}}
\newcommand{\fantomm}{{\sc FaNTOmM}}
\newcommand{\Ha}{H$\alpha$}
\newcommand{\Nii}{{\sc [Nii]}}
\newcommand{\Hi}{{\sc Hi}}

\def\kms{$\mbox{km s}^{-1}$}
\def\kmskpc{$\mbox{km s}^{-1}\mbox{ kpc}^{-1}$}

\def\deg{^\circ}

\def\farcs{\hbox{$.\!\!^{\prime\prime}$}}
\def\arcs{\hbox{$\,\,\!\!^{\prime\prime}$}}

\title[Bar pattern speed and position of the circumnuclear ring in NGC\,1097]
{Bar pattern speed and position of the circumnuclear ring in NGC\,1097}

\author[Pi\~nol-Ferrer et al.]{N. Pi\~nol-Ferrer$^{1,2}$\thanks{e-mail: npi@astro.su.se}, 
K. Fathi$^{1,2}$, 
C. Carignan$^{3,4}$, 
J. Font$^{5}$, 
O. Hernandez$^{4}$, \newauthor
R. Karlsson$^{1}$, 
G. van de Ven$^{6}$  \\ \ \\
$^{1}$Stockholm Observatory, Department of Astronomy, Stockholm University, AlbaNova Centre, 106 91 Stockholm, Sweden\\
$^{2}$Oskar Klein Centre for Cosmoparticle Physics, Stockholm University, 106 91 Stockholm, Sweden\\
$^{3}$Department of Astronomy, University of Cape Town, Private Bag X3, Rondenbosch 7701, South Africa\\
$^{4}$D\'epartement de physique, Universit\'e de Montr\'eal C.P. 6128, succ. Centre-Ville, Montr\'eal, H3C 3J7, Canada\\
$^{5}$Instituto de Astrof\'\i sica de Canarias, c/V\'\i a L\'actea, s/n, 38 205, La Laguna, Spain\\ 
$^{6}$Max Planck Institute for Astronomy, Knigstuhl 17, 69117, Heidelberg, Germany
}
\begin{document}

\pagerange{\pageref{firstpage}--\pageref{lastpage}} \pubyear{2010}

\maketitle

\label{firstpage}

\begin{abstract}
We present the first galactic-scale model of the gas dynamics of the prototype barred Seyfert~1 galaxy NGC\,1097. We use large scale \fantomm\ Fabry-Perot interferometric data covering the entire galactic disc and combine the distribution and kinematics maps with high resolution two-dimensional spectroscopy from the Gemini telescope. We build a dynamical model for the gravitational potential by applying the analytic solution to the equations of motion, within the epicyclic approximation. Our model reproduces all the significant kinematic and structural signatures of this galaxy. We find that the primary bar is 7.9$\pm0.6$~kpc long and has a pattern speed of $36\pm2$~\kmskpc. This places the corotation radius at 8.6$\pm$0.5~kpc, the outer Lindblad resonance at 14.9$\pm$0.9~kpc and two inner Lindblad resonances at 60$\pm$5~pc and 2.9$\pm$0.1~kpc. These derivations lead to a ratio of the corotation radius over bar length of 1.0--1.2, which is in agreement with the predictions of simulations for fast galaxy bars. Our model presents evidence that the circumnuclear ring in this galaxy is not located near any of the resonance radii in this galaxy. The ring might have once formed at the outer inner Lindblad resonance radius, and it has been migrating inward, toward the centre of the galactic gravitational potential.

\end{abstract}


\section{INTRODUCTION}
\label{intro}

The detailed understanding of a galaxy's gravitational potential is imperative for setting a realistic and physical scenario for its formation and evolution. The gravitational potential is the sole actor in driving the initial gravitational collapse during a galaxy's formation phase followed by possible follow-up interactions and mergers with neighbouring galaxies. Moreover, the evolution of structures in any galaxy is primarily governed by the strength and shape of its gravitational potential. Environmental effects may create evolution of structures by means of gravitational interactions, though, when structures evolve according to secular evolution scenarios, it is primarily governed by the strength and shape of the galactic gravitational potential.

One galaxy that has proven to be of particular importance for studying a wide range of gravitational effects is the nearby Seyfert~1 barred spiral galaxy NGC\,1097. This galaxy is interesting since it displays the presence of a number of  morphological and kinematic features, that are all interlinked by its gravitational field. At $\sim1 $~kpc, an almost circular ring-like feature marks a remarkable transition between the prominent $R\sim~8$~kpc galactic bar and the smooth region interior to the circumnuclear ring \citep{sersic58}, placed at the turnover radius of the rotational curve \citep{1979NZJS...22..325W}. The bar hosts two prominent dust lanes, both originated at the outer edges of the bar, that cut the inner ring, where nuclear spiral arms continue down to $\sim3$~pc distance from the active nucleus \citep{Fathi2006}, probably being the continuation of the large scale dust lanes.% giving the impression of continuity between large and nuclear scales}. 

\begin{table}
 \caption{Physical parameters adopted througout this work.}
   \label{tab:1097}
    \centering   
      \begin{threeparttable}
%\scalebox{4}{
         \begin{tabularx}{\linewidth}{|X|X|}
%         \toprule
	\hline
	\multicolumn{2}{c}{NGC\,1097} \\
	\hline
%             \midrule
	Coordinates & RA: 02:46:19  \, \,\,Dec: -30:16:30\\
	 \gray Morphology & SB(s)b\\		
	 Type & Seyfert~1\\
	\gray Distance & 14.5~Mpc$^a$\\
	P.A. & 126--131$^{\circ \, b}$\\
	\gray inclination & 35$^{\circ \,c}$ \\
	Bar radius & 107\arcs$^d$\\
%         \bottomrule
	\hline
       \end{tabularx}%}
    \begin{tablenotes}
    \item[a] \citet{1988ngc..book.....T}, \item[b] \citet{1979NZJS...22..325W,Fathi2006,2007ApJ...671.1388D,2009ApJ...696..448H,vandeVenFathi2010},
    \item[c] \citet{Fathi2006},
 \item[d] \citet{2004A&A...415..941E}
%$^a$ \citet{2004A&A...415..941E}, $^b$ \citet{1988ngc..book.....T}
    \end{tablenotes}
 \end{threeparttable}
\end{table}

While these overall features make this galaxy interesting for the studies of the evolution of structures in barred spiral galaxies, the discovery of broad ($\sim 10000$ km/s) double-peaked H$\alpha$ emission lines by \citet{1993ApJ...410L..11S} also makes NGC\,1097 an ideal laboratory for studying the `fate' of the gas accumulated in the centres of active galaxies \citep[e.g., ][]{StorchiBergmann2003}. NGC\,1097 is thus most suitable for studying the processes that cause the material/fuel to lose its angular momentum and fall from the outer galactic edge toward the galactic centre. In rotating systems, perturbations can cause the potential to become non-axisymmetric, and torques exerted by the subsequent non-axisymmetric features are able to drive material toward the centre of their host galaxy (e.g., \citealt{1984MNRAS.209...93S}, \citealt{Shlosman1989}). However, some authors as \citet{1990A&A...236..333H}, \citet{2001sac..conf..223C} or \citet{2005A&A...441.1011G} suggested that, although it is straightforward to transport gas down to the central few hundreds of parsec and induce enhanced star formation, it is more difficult to make the gas reach smaller scales required to fuel an active galactic nucleus (AGN). 

The somewhat enhanced frequency of nuclear spirals at the centre of active galaxies as compared to non-active galaxies (e.g., \citealt{Martini2003}), supports the hypothesis that nuclear spirals may aid in finalizing the last leg of the journey of the gas onto an AGN. In NGC\,1097, nuclear spirals were found in images by \cite{Lou2001}, and kinematically confirmed by \cite{Fathi2006}. Later, \cite{vandeVenFathi2010}, \cite{Davies2009} and \cite{pinol2011} measured the inflow rates of multiple phases of the interstellar medium along the nuclear spiral arms. However, none of these studies have appropriately accounted for the galactic-scale gravitational potential, simply due to the lack of observational data, what makes impossible to completely understand the nature of these spiral arms in NGC\,1097. 

Notwithstanding, it is  imperative that one builds a global and realistic model for the overall galactic-scale gravitational potential in order to explore the physical processes responsible for the nuclear spirals and that act on the mass transfer along them. Such model has been missing to date.

Here we present the global dynamical model of NGC\,1097, based on a comprehensive set of high-resolution imaging and kinematic data across the face of the galaxy. The model is based on the analytic solution of the equations of motion within the epicyclic approximation, in which we introduce a gravitational potential derived from a two-dimensional ionized gas velocity field and an infrared image. Our model simultaneously accounts for the galactic bar and spiral arms, and we have introduced a damping coefficient for adequate appearance of the resonance radii in such a way that the model can reproduce the data. The analytic solution of the equations of motion describes the response of interstellar matter, originally in circular orbits, to the gravitational potential. These solutions can be computed in few seconds, making this methodology/technique very efficient to study the large parameter space involved. %in just a few seconds, which makes it an efficient code for surveying a large parameter space.

\section{DATA}
\label{sec:data}

The dynamical model presented here is based on two-dimensional kinematic measurements of the \Ha, \Nii\ and \Hi\ data. 

\begin{figure*}
   \centering
\includegraphics[width=0.99\linewidth, trim=8mm 3mm 0mm 0mm]{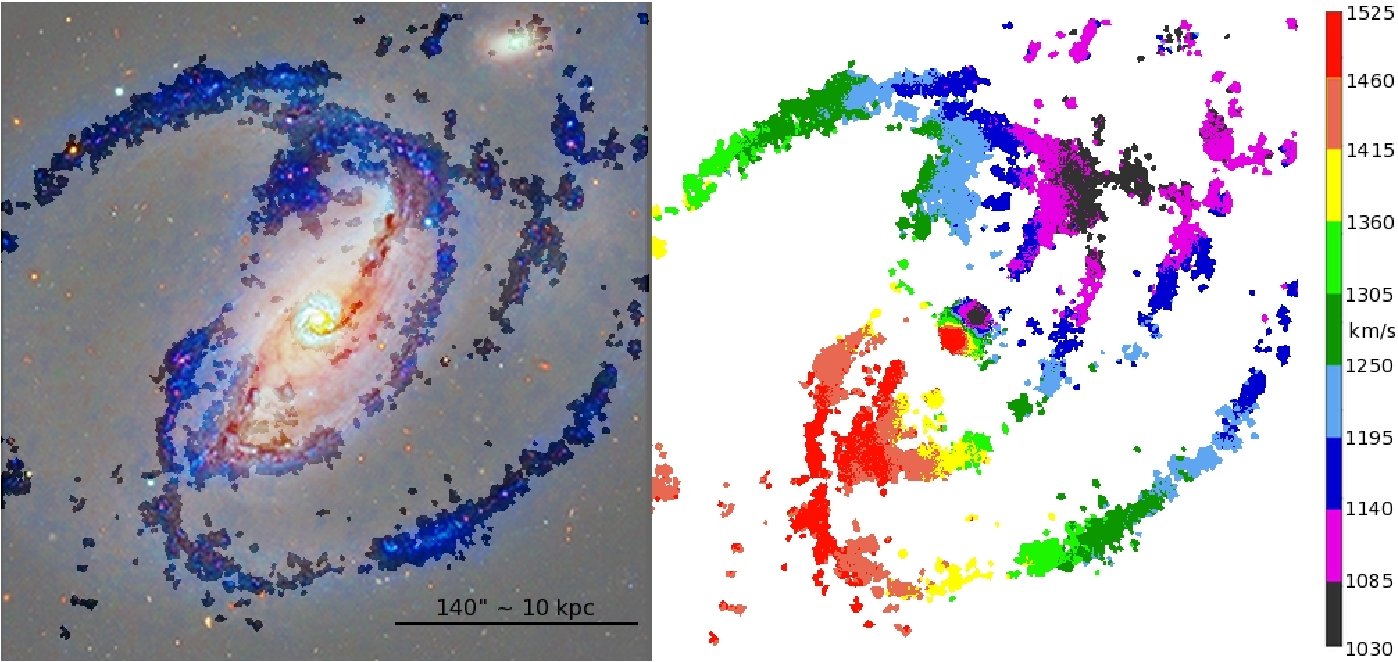}
\caption{Left: Composite three-colour image of NGC\,1097 taken with the VIMOS instrument on the 8.2-m Melipal (Unit Telescope 3) of ESO’s Very Large Telescope (Credit: ESO/Gendler) with the contour map of the FaNTOmM Fabry-Perot observations. Pixels with Signal-to-Noise larger than 5 are highlighted. Right: H$\alpha$ velocity field at a spatial sampling of 0.83\arcsec, taken with \fantomm\ on the ESO La Silla 3,6m. In both panels, north is up and east is left. The GMOS-IFU data used throughout this paper covers a region interior to the central starforming ring.}
\label{fig:veldata}
\end{figure*}

The \Ha\ kinematics and distribution are products of the \fantomm\ Fabry-Perot interferometric observations at the 3.6~m telescope from the European Southern Observatory, La Silla, Chile. Instrument specifications can be found in \citet[]{Hernandez2003}, and a preliminary presentation of the NGC\,1097 data can be found in \citet{Dicaire2008}. The Fabry-Perot data are a mosaic that cover the whole galaxy, with a spatial resolution of 1\farcs9 and the spectral resolution 15~\kms\ (see \mbox{Fig.~\ref{fig:veldata}-\emph{right panel}}). %{\bf(remember that spatial resolution and spatial sampling are not the same. pixel size is not resolution.)}.

To get a more detailed view of the central kpc, we use two-dimensional \Ha\ and \Nii\ velocity fields, with 0\farcs3 spatial resolution and 85~\kms\ spectral resolution, obtained with the Gemini South Telescope's Integral Field Unit (GMOS-IFU). The data and all corresponding quality and reduction specifications can be found in \citet{Fathi2006}. The velocity information of both lines are almost identical with equal rotation curves, with the only difference that the \Ha\ velocity field appears slightly noisier. For illustration purposes, when presenting the central velocity field, we will therefore display the \Nii\ velocity field throughout this paper.

In order to compare the kinematics in the outer parts of the galaxy, we use the \Hi\ velocity field presented by \citet{Higdon2003}. This velocity field was observed with VLA in the hybrid DnC configuration, with a channel separation of 20.7~\kms\ and a synthesized beam FWHM of 56\arcs.%a spatial sampling of 10\arcs.

%The \Hi\ neutral gas rotation curve was imported from \citet{1999ApJ...523..136S}, which we cross-checked with the \fantomm\ velocity field covering the outer parts of this galaxy. This consistency check was made to ensure that the somewhat sparsely covered outer parts of the rotation curve are not suffering from systematic effects due to relatively poor sampling of the galactic rotation curves. The \Hi\ data confirms the velocity measurements that we have derived from the ionized gas
%data have been presented in {\bf van der Hulst, J.M. et al. (1970-1999 check the year)}, which velocity field we cross-checked with the \fantomm\ velocity field covering the outer parts of this galaxy. This consistency check was made to ensure that the somewhat sparsely covered outer parts of the rotation curve are not suffering from systematic effects due to relatively poor sampling of the galactic rotation curves. the \Hi\ data confirm the velocity measurements that we have derived from the \Ha\ emission line.

Finally, to probe the overall galactic potential, we further use imaging data from the old stellar population based on the 3.6 $\mu$m image from the Spitzer telescope \citep{Kennicutt2003}. The Spitzer observations cover the galaxy with a spatial resolution of 1\farcs7.%at a pixel size of 0\farcs75.

%\begin{figure}
%   \centering
%\includegraphics[width=0.3\linewidth, trim=20mm 0mm 18mm 0mm]{figs/rv_gmos.jpg}
%\includegraphics[width=0.3\linewidth, trim=29mm 0mm 9mm 0mm]{figs/rv_gmosconv.jpg}
%\includegraphics[width=0.3\linewidth, trim=38mm 0mm 0mm 0mm]{figs/rv_fantg.jpg}\\
%\includegraphics[width=0.3\linewidth, trim=20mm 0mm 18mm 0mm]{figs/in_gmos.jpg}
%\includegraphics[width=0.3\linewidth, trim=29mm 0mm 9mm 0mm]{figs/in_gmosconv.jpg}
%\includegraphics[width=0.3\linewidth, trim=38mm 0mm 0mm 0mm]{figs/in_fantg.jpg}
%\caption{\emph{Upper row:} GMOS-IFU velocity field as observed, binned and convolved to match the resolution of the \fantomm\ velocity field covering the same region. {\em Lower row:} same as the upper row, but here we illustrate the comparison between the \Ha\ distribution, in arbitrary units. Note that arbitrary units are chosen due to the the fact that this particular Fabry-Perot data cube cannot be flux calibrated.}
%\label{fig:gmosfantom}
%\end{figure}

\section{A DYNAMICAL MODEL FOR NGC\,1097}
\label{sec:3}

To build a dynamical model for the gravitational potential of NGC\,1097, we apply the analytic solution within the epicyclic approximation. The analytic model we use was first developed by \citet{1976ApJ...209...53S}, and then by \citet{Lindblad1994} and \citet{Wada1994} whom included a special analysis on the resonant radii. This analysis was used to fit observations by \citet{Lindblad1996}, \citet{1999ApJS..124..403S}, \citet{2000PhDT.........6B} or \citet{2007A&A...471..113B}, but assuming always a barred gravitational potential. For NGC\,1097, we generalize this analytic model for an arbitrary galaxy potential.

%The analytic model we use is a generalization for an arbitrary galaxy potential, which was first developed by \citet{1976ApJ...209...53S} and then by \citet{Lindblad1994} and \citet{Wada1994} with a special analysis on the resonant radii. This analytical model has been used to fit observations by \citet{Lindblad1996}, \citet{1999ApJS..124..403S} or \citet{2007A&A...471..113B}, but not using an arbitrary gravitational potential. 

Hence, we introduce the arbitrary gravitational potential in the analytic study and adjust the appearance of the resonances by using damping coefficients. The generalized method is thoroughly described and extensively tested by \citet[][, \Epic]{PinolFerrer2012}, even using the strong barred galaxy NGC\,1365. Accordingly, we calculate gas orbits and their velocity and density distribution for an arbitrary gravitational potential. 

\Epic\ solves the equations of motion in a solid body rotating, time-independent galactic potential. We assume that the potential is rotating with a pattern speed $\Omega_{\rm p}$, and we study the system in a corotating frame. We express this arbitrary gravitational potential in polar coordinates, dividing it between an axisymmetric component and a non-axisymmetric perturbation:%as in equation (\ref{eq:potentialap}), where $\vartheta_m$ is the spiral phase. 

\begin{equation}
\Phi(r,\theta)=\Phi_0(r)+\Phi_1(r,\theta).
\label{eq:potentialap}
\end{equation}

The equations of motion are solved assuming the first-order epicyclic approximation (see also \citealt{BinneyTremaine2008}, p.189). Once linearised, we introduce a frictional force proportional to the deviation from circular motion with a damping coefficient, $2\lambda$ (see \citealt{Wada1994}, \citealt{Lindblad1994}), for adequate appearance of the resonances

Then, we introduce $\xi(t)$ and $\eta(t)$ as deviations from circular motion, ($r_0, \theta_0$),

\begin{align}
\centering
r&=r_0+\xi \\
\theta&=\theta_0 + (\Omega - \Omega_{\rm p}) \,t + \frac{1}{r_0} \eta,
\end{align}

\noindent
and write the full solution of the linearised equations as:

\begin{align}
\nonumber \xi &= c e^{-\alpha\lambda t} \cos \kappa(t-t_0) \\ 
\   &+ \sum_{m=1}^{n} \big[ d_m \cos m(\theta-\vartheta_m) + e_m \sin m(\theta -\vartheta_m) \big], \label{eq:xisolap}\\
\nonumber \eta&= - \frac{2\Omega}{\kappa}c e^{-\beta\lambda t} \sin \kappa(t-t_0)  \\
\   &+ \sum_{m=1}^{n} \big[ g_m \sin m(\theta-\vartheta_m) + f_m \cos m(\theta -\vartheta_m) \big],\label{eq:etasolap}
\end{align}

\noindent
where $\vartheta_m$ is the spiral phase, $\Omega$ the circular angular velocity at radius $r_0$, $\kappa$ is the epyciclic frequency which can be expressed as $\kappa^2=4\Omega^2-4\Omega A$, $A$ is the Oort constant, $\alpha$ and $\beta$ are functions of $\Omega$ and $\kappa$, $c$ is an arbitrary constant and $d_m$, $e_m$, $g_m$ and $f_m $ are given in \citet{PinolFerrer2012}. The second terms on the left side of eqs. (\ref{eq:xisolap}) and (\ref{eq:etasolap}) give the forced oscillation due to the perturbing force. The first terms give the damped oscillation with the epicyclic frequency $\kappa$ around these guiding centra. 

With these solutions, \Epic\ computes the deformation of the initially circular orbits of the guiding centre, the velocity fields and the structure created by the arbitrary gravitational potential. Hence, the code pictures a steady state scenario produced by a weak perturbing potential. All details regarding the mathematical solution can be found in \cite{PinolFerrer2012}.

The case of NGC\,1097 presents a strong bar with a strength, $|\Phi_1|_{\rm min}/|\Phi_0|_{\rm min}$,
 of 0.02. As in the case of NGC~1365 analysed in \citet{PinolFerrer2012} and which strength is 0.01, although NGC\,1097 possesses a strong perturbation, \Epic\ is capable of giving us an estimation of important parameters such as the bar pattern speed and resonance radii by fitting the results to observational data. However, a model based on numerical simulations that accounts for strong gravitational  perturbation is needed in order to fully reproduce the physical scenario at play.

\subsection{Fourier Decomposition of the Velocity Field}
\label{sec:kinematics}

We decompose the observed Fabry-Perot velocity field into Fourier terms up to and including the third order. Numerous previous authors have shown that this analysis is particularly useful since every component of the velocity Fourier series will bring information about the  underlying gravitational potential. The $m$ mode of the perturbing potential will give rise to the $m-1$ and $m+1$ Fourier terms of line of sight velocity components (e.g. \citealt{Canzian1993,Schoenmakers1997,2003SSRv..105....1F}). Further discussion on this point can be found in the appendices of  \cite{vandeVenFathi2010} and \cite{PinolFerrer2012}.

%{\bf , and Fridman \& Khoruzhii 2003 Space Science Review 105, 1. Also of there is a paper usig the tool called ResWri, part of the Gipsy package).} 

The method we use for the decomposition is similar to the one described in \citet{Fathi2005}, which we only explain briefly here. We divide the velocity field into concentric elliptical annuli, which major axis is defined by $a_{i+1}=a_i+a_i^\gamma dr$, where $a_i$ is the major axis of each ring, $dr$ a fixed width of 3 pixels and $\gamma = 0.5$. As we expect that rotation is dominant, we start by first only fitting a rotating disc and make a $\chi^2$ analysis to obtain the inclination $i$, systemic velocity, position angle P.A., and the central coordinates $(x_0, y_0)$ of the disc. After fixing these parameters, we make the final $\chi^2$ fit to the desired modes (up to and including 3rd order in our case) of the Fourier series decomposition shown in eq.~(\ref{eq:vlos}), where $V_{\rm los}$ is the line of sight velocity and $V_{\rm sys}$ the systemic velocity. We are assuming in this method that every pixel in the velocity field corresponds to only one position in the inclined disc, and we use a Monte-Carlo Bootstrap method for the estimation of the errors of the Fourier series components, considering in this estimation also the error from the observations.

\begin{equation}
V_{\rm los} = V_{\rm sys} + \sum_{n=1}^k [c_n(r) \cos n\theta + s_n(r) \sin n\theta] \sin i
\label{eq:vlos}
\end{equation}

\begin{figure*}
   \centering
\includegraphics[width=0.32\linewidth, trim=20mm 0mm 5mm 0mm, angle=0]{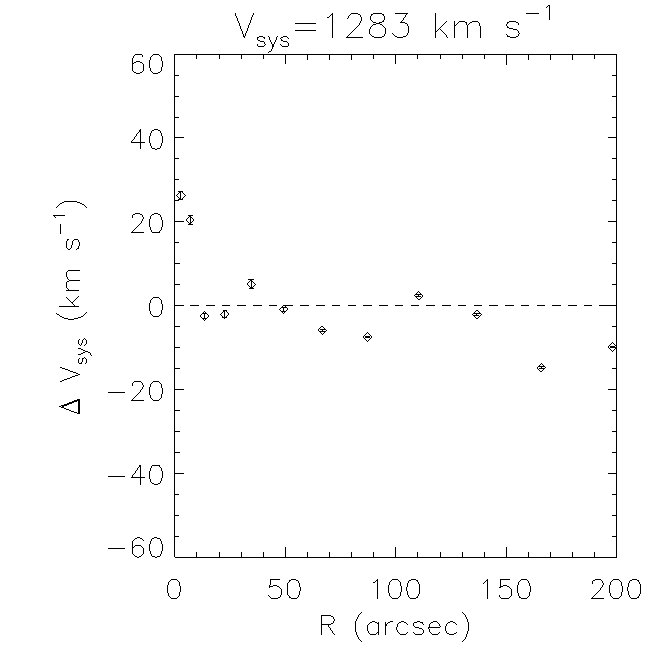}%\\ \vspace{0.4cm}
\includegraphics[width=0.32\linewidth, trim=12.5mm 0mm 12.5mm 0mm, angle=0]{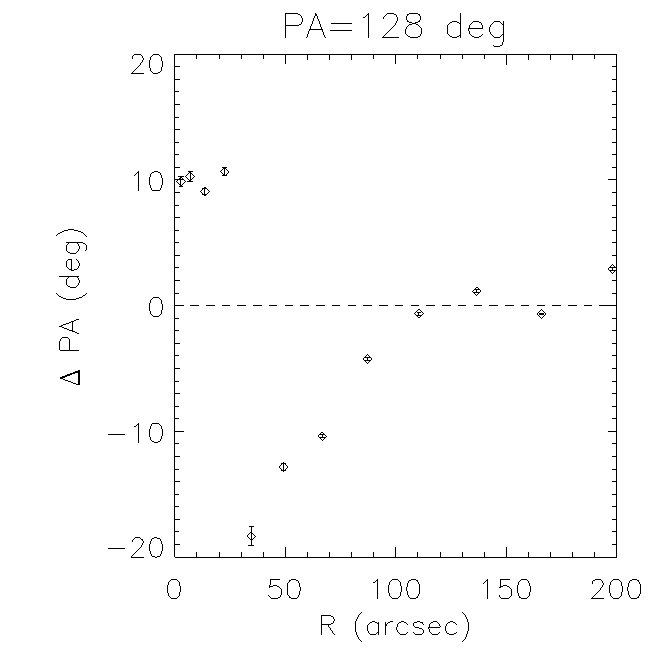} %\\ \vspace{0.4cm}
\includegraphics[width=0.32\linewidth, trim=5mm 0mm 20mm 0mm, angle=0]{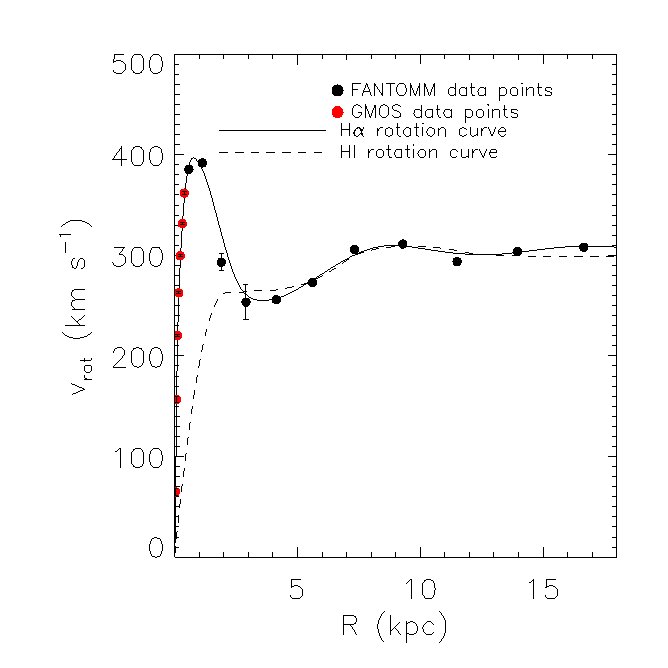} \\ \vspace{0.4cm}
\includegraphics[width=0.32\linewidth, trim=20mm 0mm 5mm 0mm, angle=0]{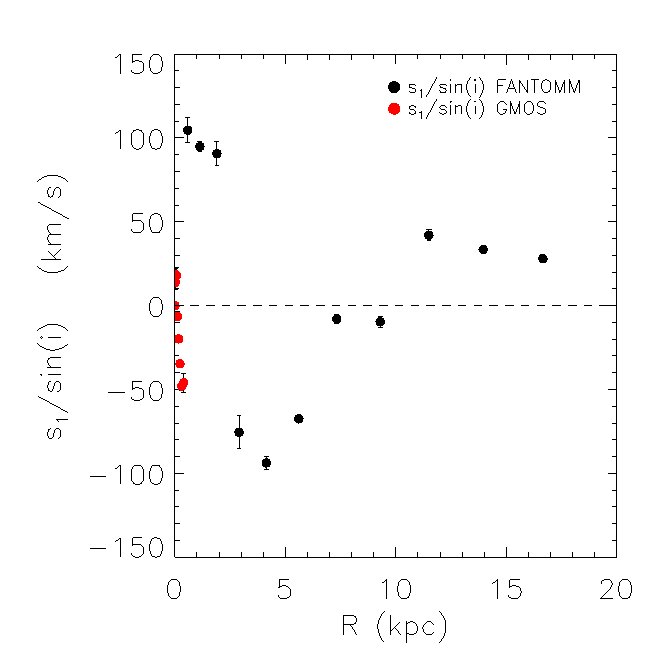}%\\ \vspace{0.4cm}
\includegraphics[width=0.32\linewidth, trim=12.5mm 0mm 12.5mm 0mm, angle=0]{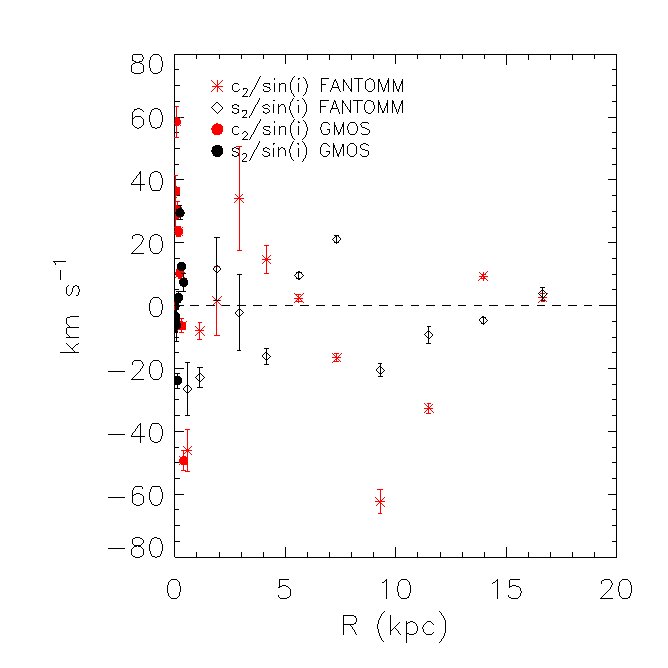}% \\ \vspace{0.4cm}
\includegraphics[width=0.32\linewidth, trim=5mm 0mm 20mm 0mm, angle=0]{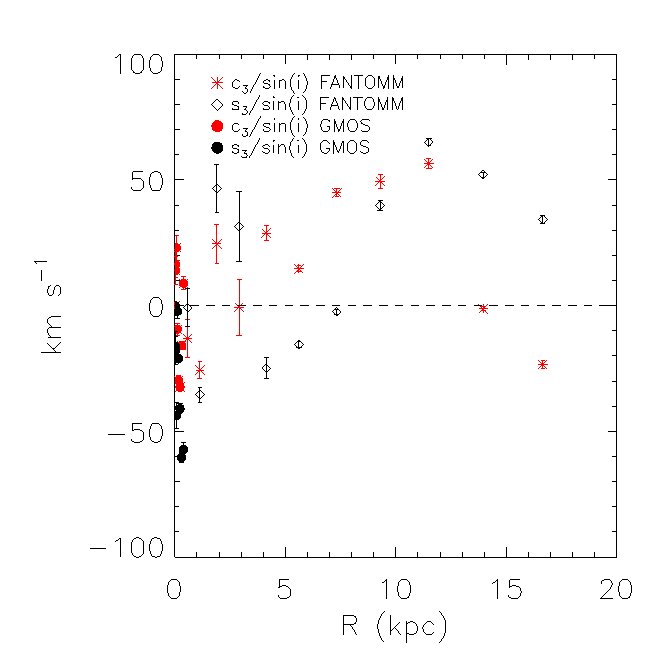} \\ \vspace{0.4cm}
\includegraphics[width=0.32\linewidth, trim=20mm 0mm 5mm 0mm, angle=0]{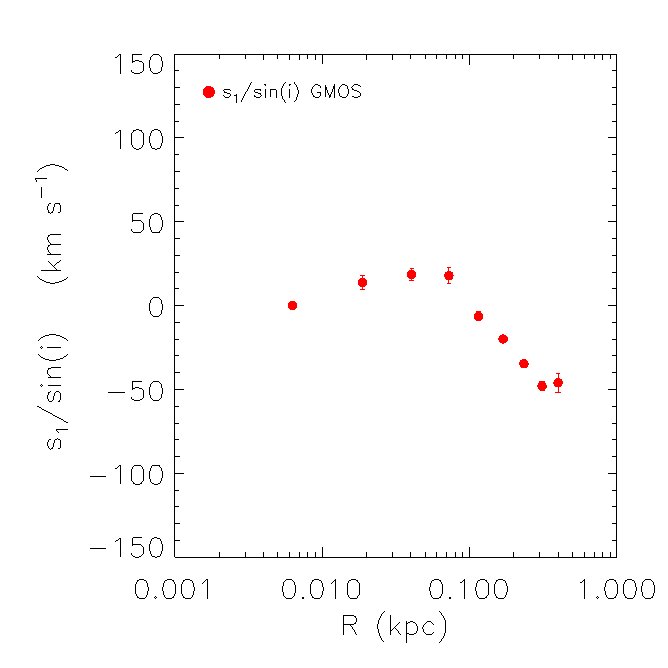}
\includegraphics[width=0.32\linewidth, trim=12.5mm 0mm 12.5mm 0mm, angle=0]{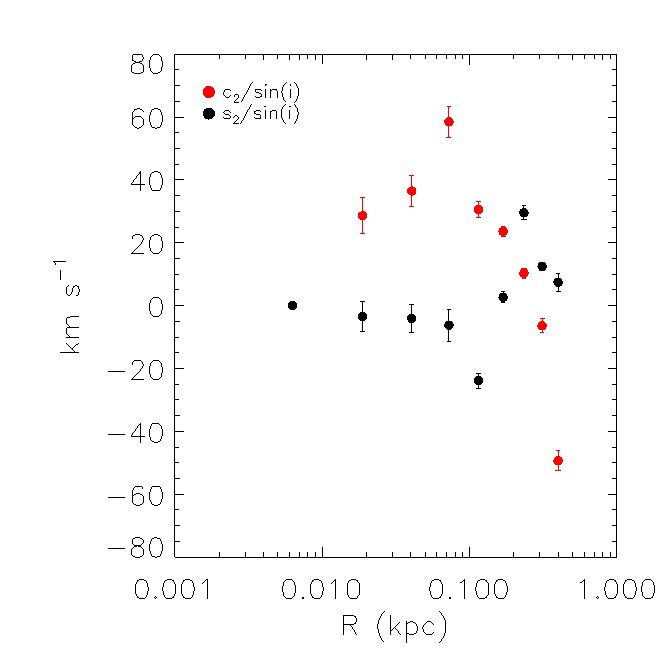}
\includegraphics[width=0.32\linewidth, trim=5mm 0mm 20mm 0mm, angle=0]{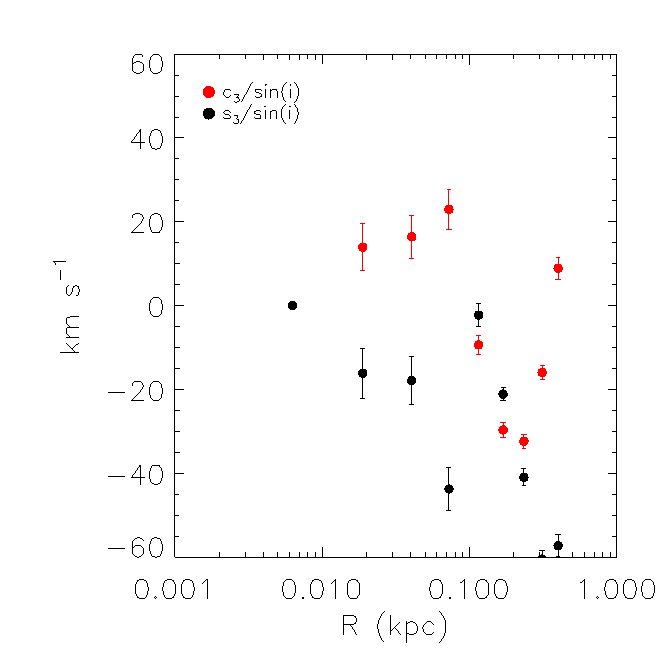}
\caption{Fourier expansion for the high-resolution ionized gas velocity field of NGC\,1097 up to the third mode. We have combined the components derived from the nuclear region (GMOS-IFU data) and from the large scale galaxy (\fantomm\ data). We fix the systematic velocity and the kinematic position angle in the analysis of \fantomm\ data as shown in the two first upper graphs. In the third upper graph we show the derived rotation curve as a combination of the two data sets GMOS-IFU and \fantomm\ together with the \Hi\ rotation curve. In the two lower rows, we present the different Fourier terms with linear and logarithmic $x$-axes.%\bf this figure is not very easy to read. you can mix the c2 and s2 in the same panel with different symbols, and same for s3 and c3. Then you can decrease the space between the panels to make the plots a bit bigger,  and all in one row, then in a second row, plot everything again but in logarithmic x-axis (/xlog command in IDL).
}
\label{fig:rot}
\end{figure*}

\begin{figure*}
   \centering
\includegraphics[width=0.49\linewidth, trim=15mm 0mm 13mm 0mm, angle=0]{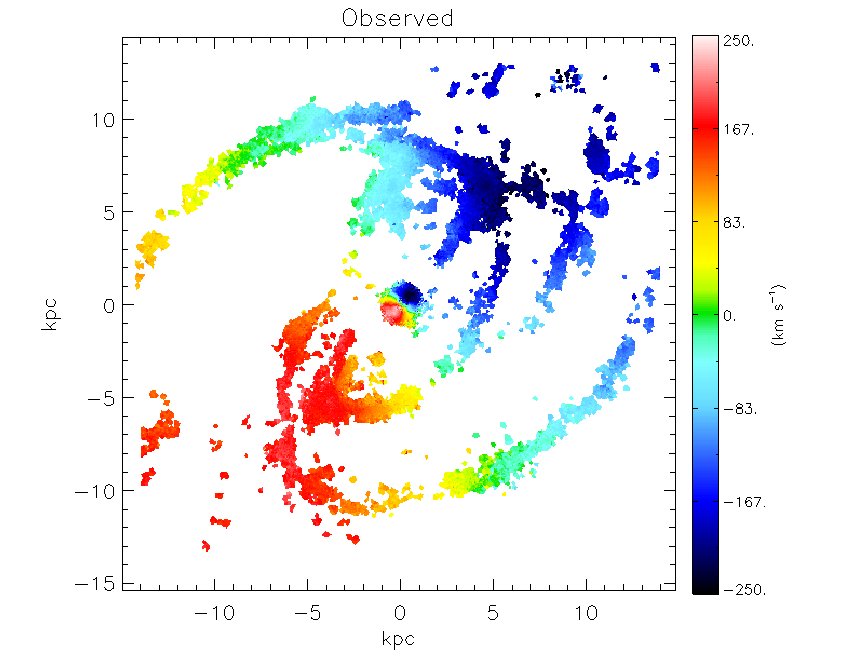}
\includegraphics[width=0.49\linewidth, trim=13mm 0mm 15mm 0mm, angle=0]{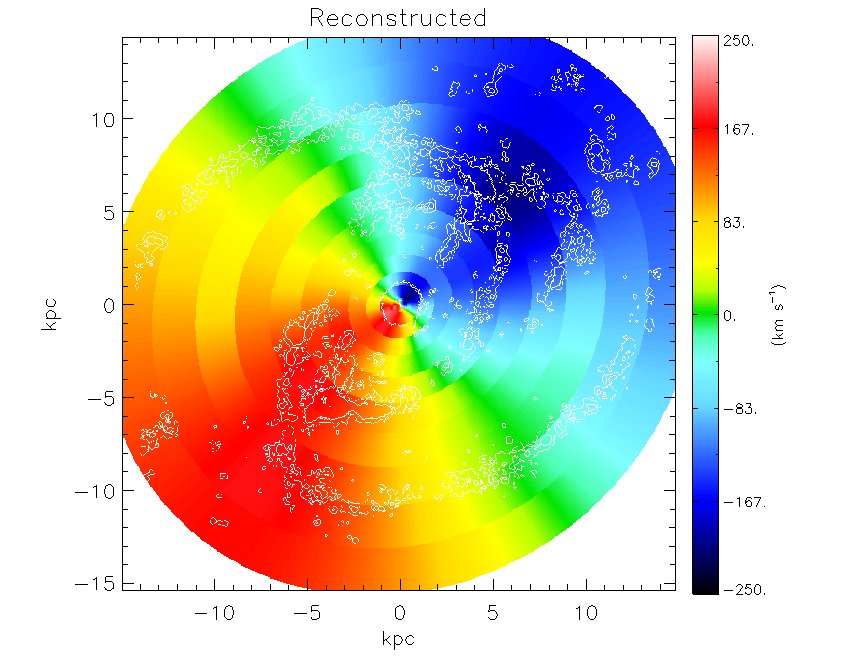} \\ \vspace{0.4cm}
\includegraphics[width=0.49\linewidth, trim=15mm 0mm 13mm 0mm, angle=0]{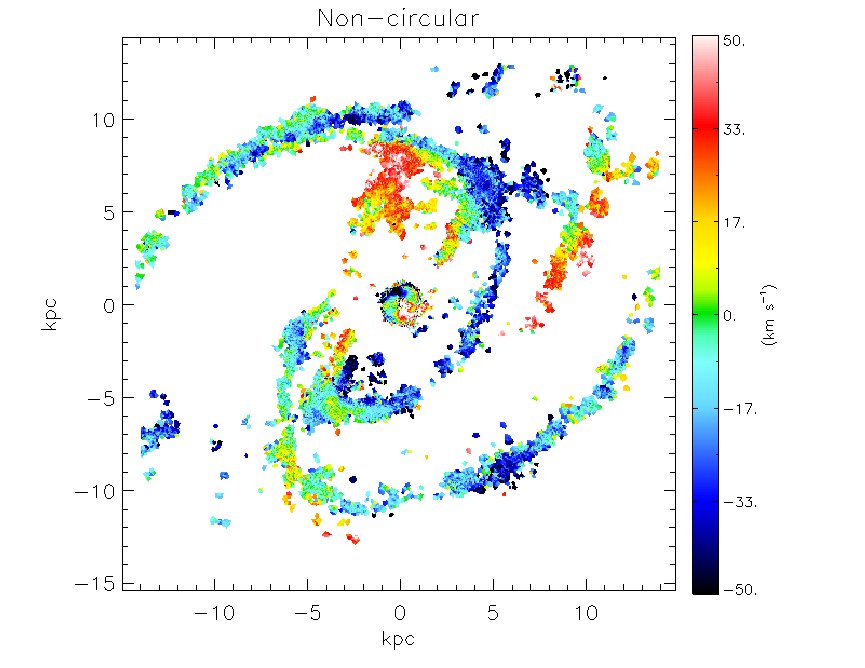}
\includegraphics[width=0.49\linewidth, trim=13mm 0mm 15mm 0mm, angle=0]{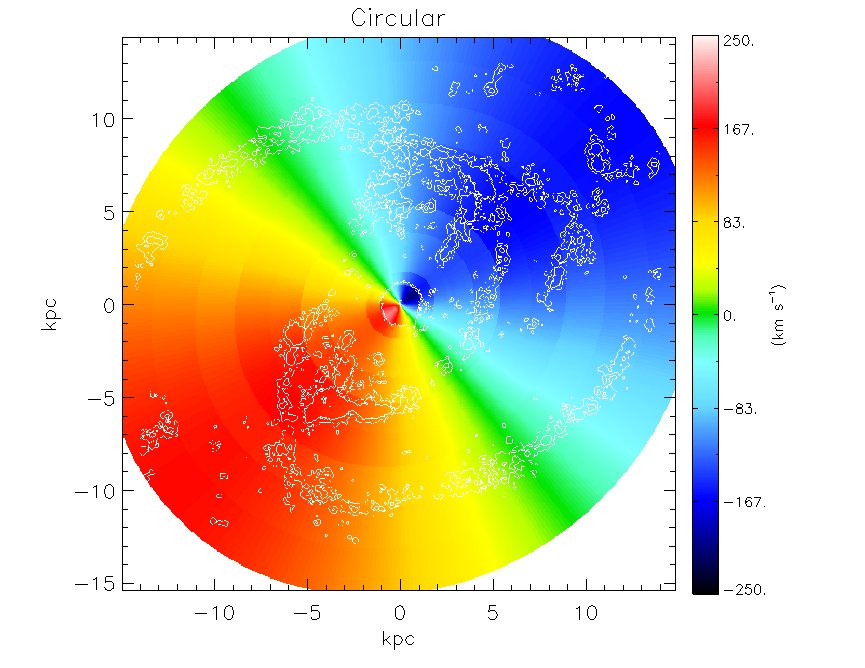}
\caption{\emph{Upper left:} Observed Fabry-Perot \Ha\ velocity field. \emph{Upper right :} Reconstructed velocity field from the first three modes of the Fourier decomposition shown in Fig.~\ref{fig:rot} (combining both the \fantomm\ and the GMOS-IFU data sets). \emph{Lower left:} Non-circular velocities that remain after removing the rotating component from the observed velocity field; and \emph{Lower right:} a rotating disc model fitted to the observed velocity field (combining both the \fantomm\ and the GMOS-IFU data sets).}
\label{fig:FHDmap}
\end{figure*}

We obtain the Fourier decomposition of both the GMOS-IFU and the \fantomm\ velocity maps applying this procedure. For the GMOS-IFU data, \citet{vandeVenFathi2010} found almost identical parameters. This type of analysis of the entire galaxy-scale of velocity field of NGC\,1097 is unprecedented. We find an average kinematic position angle (PA) for the whole galaxy, using \fantomm\ observations, of $128\pm10\deg$, and a P.A. at the nuclear kpc, from GMOS observations, of $141\pm8\deg$. This nuclear P.A. is also observed in the fit of the position angle along radius using the \fantomm\ data, see \mbox{Fig.~\ref{fig:rot}-\emph{middle top panel}}. We obtain a systematic velocity from the large scale analysis of 1283$\pm$10~\kms, while we find 1197$\pm$5~\kms for the nuclear kpc, using the GMOS-IFU data.  These discrepancies may be produced by the prominent non-circular motions in the inner region or the lopsided instabilities in the outer disc of the galaxy. However, NGC\,1097 cannot be clearly classified as a lopsided galaxy since an estimation of a $A_1$ parameter, which is an indicator of this kind of instability \citep{2005A&A...438..507B}, is not significant. Therefore, this highlight the importance of using kinematic data that cover the whole galaxy, since the inner regions present complications in the derivation of these overall galactic parameters. In Fig.~\ref{fig:FHDmap} we illustrate the comparison between the observations and the recovered map from the rotating component and the full Fourier decomposition.

In addition, we obtain the Fourier decomposition of the \Hi\ velocity field in order to cross-check with the \fantomm\ velocity field at large radii. This consistency check was made to ensure that the somewhat sparsely covered outer parts of the rotation curve are not suffering from systematic effects due to relatively poor sampling of the galactic rotation curves. The derived \Hi\ rotation curve confirms the velocity measurements that we have derived from the \Ha\ emission line, \mbox{Fig.~\ref{fig:rot}-\emph{right top panel}}. The \Hi\ velocity field analysis gives an estimation of the P.A. of $127\pm5\deg$ and the systemic velocity of $1271\pm10$~\kms, which is also in concordance with the \Ha\ large scale analysis.

On the other hand, in Fig.~\ref{fig:gmosresult} is shown the non-circular velocities of the nuclear region of NGC\,1097, that is the observed velocity field with the subtraction of the circular velocities used in section~\ref{sec:pot}. As it is explained in section~\ref{sec:pot}, these circular velocities are estimated interpolating the rotational curve derived from \fantomm\ and from GMOS data, what produces a slightly different curve in the inner region from the rotation derived in \citet{vandeVenFathi2010}. This discrepancy explains the differences that can be observed between our derived non-circular velocities and their map.  

\subsection{Constructing a Potential}\label{sec:pot}

We derive the gravitational potential of NGC\,1097 assuming that it is formed by a dominating axisymmetric component, $\Phi_0$, and a weak non-axisymmetric perturbation, $\Phi_1$. We express the potential by its Fourier decomposition in polar coordinates, $(r,\theta)$:
\begin{align}
\Phi(r,\theta)=&\Phi_0(r)+\Phi_1(r,\theta)\nonumber\\
=&\Phi_0(r)-\sum_{m=1}^{n} \Psi_m(r) \, \cos m(\theta - \vartheta_m(r))
\label{eq:potential}
\end{align}
where $\Psi_m(r)$ is the coefficient of the trigonometrical series and $\vartheta_m$ is the spiral phase.

\begin{figure}
   \centering
\includegraphics[width=0.65\linewidth, trim=10mm 0mm 15mm 0mm, angle=0]{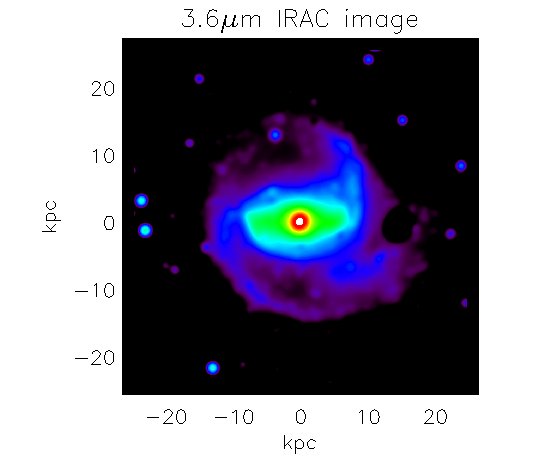}
\includegraphics[width=0.65\linewidth, trim=10mm 0mm 15mm 0mm, angle=0]{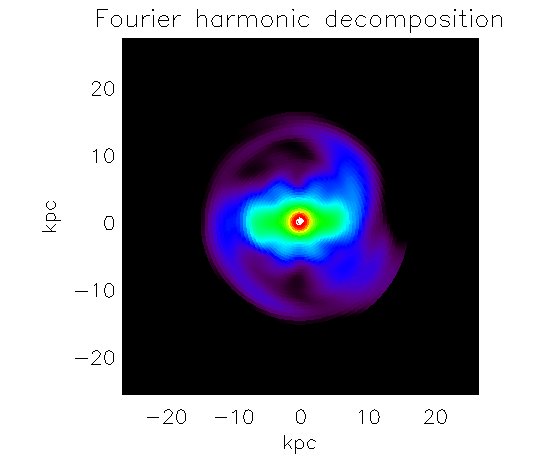}
\caption{\emph{Upper image:} 3.6~$\mu$m emission image of NGC\,1097 convolved with a two-dimensional 2~kpc sigma Gaussian. \emph{Lower image:} Fourier decomposition of the image considering $m=0$, 2, 3, 4, 5 and 6 modes.}%the six first modes but the first one. \bf what does the last part mean? "six first modes but the first one", does it mean that m=1 is missing?, then you could write "... considering $m=0$, 2, 3, 4, 5 and 6 modes"}
\label{fig:3.56image}
\end{figure}

We estimate the axisymmetric component of the potential from the rotation curve of the galaxy. This rotation curve is derived from the Fourier decomposition of the velocity field, described in section~\ref{sec:kinematics}. We combine the rotation curve from the large scale observations with the nuclear \Ha\ data. Then we interpolate the data points obtained from \fantomm\ and GMOS data sets, see \mbox{Fig.~\ref{fig:rot}-\emph{right top panel}}. We also smooth this interpolated curve and its derivative in order to get a suave result with \Epic. We cannot trust the inner 70--100~pc of the resulting rotation curve since data there present a high level of uncertainty due to dust, non-Gaussian emission lines, high velocity dispersion etc. \citep{Fathi2006}.

To derive the large scale gravitational potential perturbation we use the deprojected 3.6~$\mu$m emission image. We fix the photometric projection parameters at PA$=128\deg$ and $i=35\deg$ derived by the kinematic analysis. We rotate the galaxy placing its bar horizontally and mask the brightest background stars and the satellite from the image, see \mbox{Fig.~\ref{fig:3.56image}-\emph{left}}. Before we decompose the image into a Fourier series, we also convolve the 3.6~$\mu$m emission image with a two-dimensional 2~kpc sigma Gaussian profile, in order to obtain smoother Fourier components along radius.

When decomposing into its Fourier components, we consider the modes $m=$0, 2, 3, 4, 5 and 6. We exclude the mode $m=1$ in order to not displace the centre of gravity from the centre of the galaxy. That is because \Epic\, when solving the equation of motion, assumes that the centre of gravity is at rest. 

We convert the 3.6~$\mu$m surface brightness into surface density using a constant mass-to-light ratio throughout the disc. We analyse the observed rotational curve and how it is recovered for radii between 120\arcsec~and 180\arcsec~(8.4~kpc and 12.6~kpc, respectively, just outside the bar where the dust content is lower) using different ratios. We obtain the best match for a mass-to-light ratio of 1.4~\msun/L$_{\odot}^{3.6\mu\rm m}$.   %{these last two paragraphs are a bit random, reorder the text so that youy describe exactly the order of what you are doing. now it seems like you get the mass profile, and then you smooth, so one could wonder why you smooth in the end?}

From the Fourier components of the surface density, we derive the perturbing potential following the analysis using Bessel functions described in \citet[Ch. 2.6.2]{BinneyTremaine2008} and \citet{Lindblad1996}. We account for the disc thickness by using a normalized triangle density distribution along the vertical direction in the plane of the disc, i.e. 
\begin{equation}
F(z)=\left\{
  \begin{array}{l l}
    \frac{1}{z_0}(1-\frac{|z|}{z_0}) & \quad \text{for $|z|<z_0$}\\
    0 & \quad \text{otherwise}
  \end{array} \right.
\end{equation}
\noindent
where we assume a $z_0$ scale parameter equal to 1~kpc, the 13\% of the bar which is in concordance with a rotation velocity of 300~\kms\ \citep{2011ARA&A..49..301V}. %{\bf the last sentence is very compact, please explain a bit more, what is the vertical direction? is it the vertical direction in the plane of the disc or in the image?}

\begin{figure}
   \centering
\includegraphics[width=0.7\linewidth, trim=15mm 0mm 3mm 0mm, angle=0]{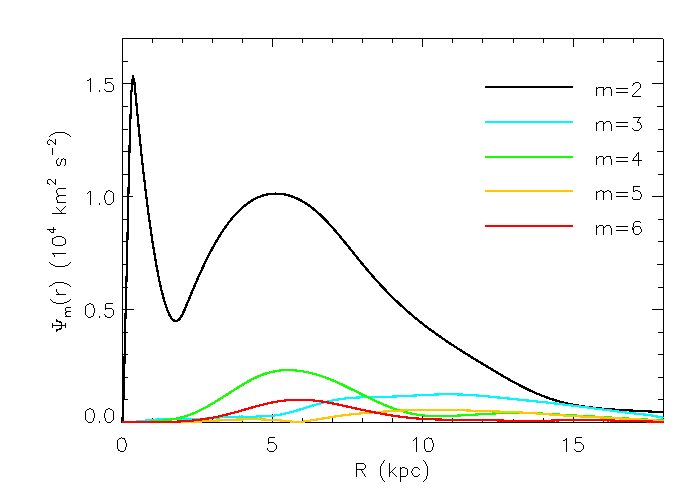}\\ \vspace{0.4cm}
\includegraphics[width=0.7\linewidth, trim=15mm 0mm 3mm 0mm, angle=0]{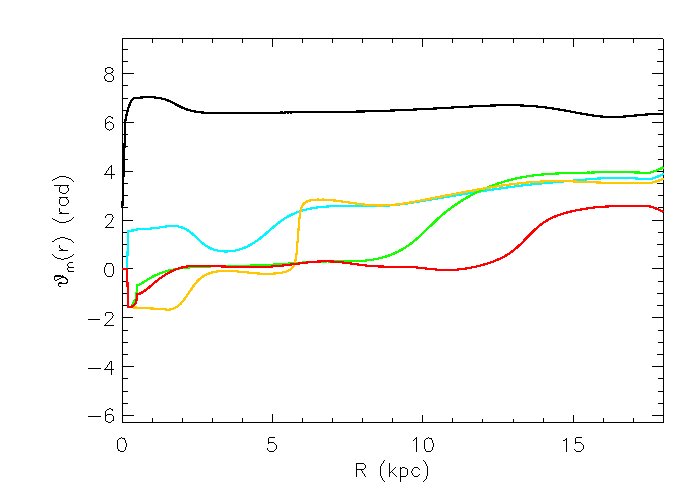}%\\ \vspace{0.4cm}
\caption{The \emph{m} modes of the perturbing potential Fourier decomposition, eq.~(\ref{eq:potential}). \emph{Upper graph:} $\Psi_m(r)$, where $m=2$ displays a strong nuclear component. Colours correspond to the same modes in both the $\Psi_m(r)$ and the $\vartheta_m(r)$ graphs. }
\label{fig:mpot}
\end{figure}

For the central 4\arcsec\ the potential was derived from the GMOS-IFU data in order to get a more accurate potential across the circumnuclear region \citep{vandeVenFathi2010}. We combine the large scale gravitational potential, for radius larger than 2~kpc, with the nuclear scale potential from GMOS-IFU data. Then, we interpolate between the nuclear potential and the large scale potential, for radii between $\sim$0.4~kpc to 2~kpc. In this transition region, the interpolation is the source of major uncertainty. We place the nuclear spiral arms at an angle that gives the best comparison between the observed non-circular velocities and those in our analytic model. %  {\bf something is worng, please re-phrase} 
We observe that, the $m=2$ mode from the GMOS data is approximately a factor 1.5 lower than the minimum of the perturbing potential derived from the Fabry-Perot data, see Fig.~\ref{fig:mpot}. That means that the derived gravitational potential in the nuclear region is much stronger that for larger radii in the galaxy. It is likely that this is a real effect, meaning that the GMOS data are more powerful in picking up the strength of the perturbing potential in the nuclear part. In Figure~\ref{fig:mpot}, $\Psi_m(r)$ is the coefficient of the trigonometric series and $\vartheta_m$ is the spiral phase, once that nuclear and large scale potentials have been combined. %The $m$ modes of the perturbing potential along radius are shown in Fig.~\ref{fig:mpot}.{\bf this sentence is not necessary, and the plot is quire ugly, so maybe you can plot is more reader-freindly.}

\subsection{\Epic: An Analytic Fit to NGC\,1097 Data}
\label{subsec:model}
We model the NGC\,1097 data, by introducing the gravitational potential estimated as described in section~\ref{sec:pot}. To finalize the model, we need to know the bar pattern speed. One way to estimate the pattern using this code is comparing the observed velocity field presented in section~\ref{sec:data}, \mbox{Fig.~\ref{fig:veldata}-\emph{right panel}}, with velocity fields generated by \Epic\ using different pattern speeds. We realize a $\chi^2$ study, varying the pattern speed and damping coefficients, fitting the velocity field generated by \Epic\, projected on the plane of the galaxy, to the observed velocity field. We use the \fantomm\ data to fit the large scale structure, using exactly the potential described in section~\ref{sec:pot}, and the GMOS data only to fit the nuclear region, using the potential derived in \citet{vandeVenFathi2010}. From the large scale fit, after a $\chi^2$ analysis for 8 different combinations of the 4 variables involved, we obtain a pattern speed of $\Omega_p=36\pm2$~\kmskpc. We estimate the error of the pattern speed considering the 1-$\sigma$ confidence level.% (black contour line in Figs.~\ref{fig:chi2}). 
%It is shown the distribution presented in the slides of the cube where the minimum $\chi^2$ is and the distribution of the collapsed cube for every pair of variables. 

Besides the potential and the pattern speed, \Epic\ needs three damping coefficients, at corotation and inner/outer Lindblad resonances, as inputs in order to generate the solution. %, see \citet{PinolFerrer2012} for further explanation of these coefficients.
 In \Epic, the damping coefficient that avoids singularities at the Lindblad resonances has been found  to have a linear dependency on the radius. However, in the case of the strong barred galaxy NGC\,1097, much larger damping coefficient at nuclear scales are needed compared with the outer ranges. For this reason, we choose the damping coefficient to be proportional to the angular frequency along radius. %In Fig.~\ref{fig:chi2}, we further observe that while there is a strong dependency of the $\chi^2$ fit on the pattern speed,  the damping coefficients are more degenerate. For this reason, 
We use a range of damping coefficients for which the gas orbits do not cross. %The nuclear scale fit gives a wide range of pattern speed values, being our previous estimation and higher values valid.

While we see in Figs.~\ref{fig:fantomresult}~and~\ref{fig:gmosresult} that \Epic\ generate approximate NGC\,1097 kinematics, we note that the non-circular velocity amplitudes are not objective since they depend on the adopted damping coefficient. Since NGC\,1097 hosts a strong bar and \Epic\ assumes a weak perturbation, high damping coefficients are needed to `dampen' the model non-circular velocity amplitudes. We think that the presence of a strong bar when assuming a weak perturbation also explains the mismatched features inside the corotation radius, the much straighter observed density ``lanes'' are not aligned by the lanes in the model and hence the observed location for the shock loci is displaced. However, other explanations such as a scenario where the spiral arms are decoupled with the bar could also contribute to some of the discrepancies between the model and observations.

%\begin{figure*}
%   \centering
%\includegraphics[width=0.99\textwidth, trim=5mm 5mm 5mm 0mm, angle=0]{../NGC1097/chi2/chi2/thmn012/chi2_clevel.jpg}\\ \vspace{0.4cm}
%\includegraphics[width=0.99\textwidth, trim=5mm 5mm 5mm 0mm, angle=0]{../NGC1097/chi2/chi2/thmn012/chi2_clevel_col.jpg}%\\ \vspace{0.4cm}
%\caption{Confidence level of the large scale $\chi^2$ study. The black contour marks the 1-$\sigma$ level. $\Omega_p$ is the pattern speed, $\lambda_0$ is the damping coefficient at the inner Lindblad resonance, $\lambda_f$ at the outer Lindblad resonance and $\varepsilon$ at corotation. The three upper plots show the $\chi^2$ distribution on the slice of the cube where the minimum value is. The three lower plots show the collapsed $\chi^2$ distribution for every pair of variables.}
%\label{fig:chi2}
%\end{figure*}

\begin{figure}
   \centering
\includegraphics[width=0.8\linewidth, trim=8mm 3mm 0mm 0mm, angle=-90]{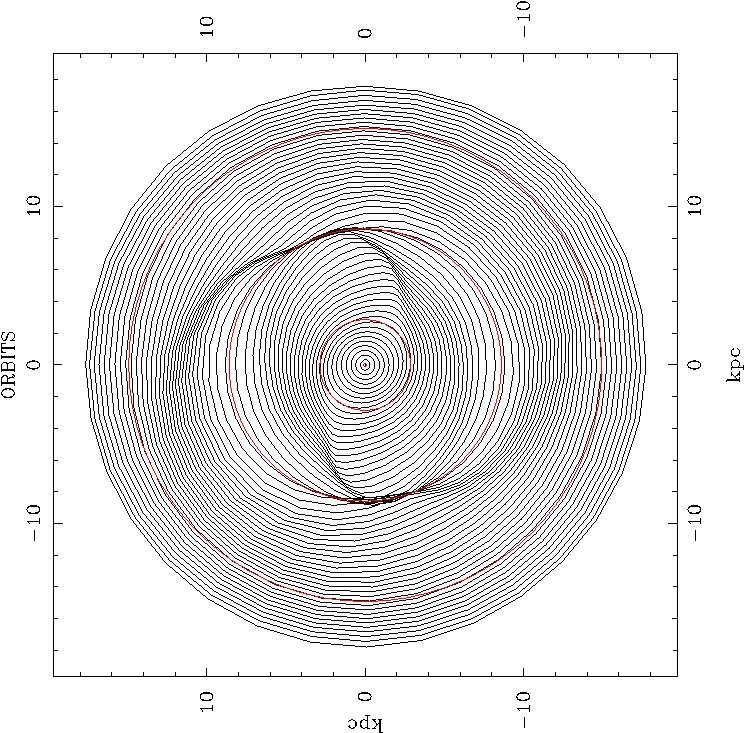}
\caption{Orbits of our model for NGC 1097 generated by \Epic\ in the plane of the disc. The bar is positioned in the horizontal axis and the radial separation between the orbits is 0.3 kpc. Red circles mark the resonance radii.}
\label{fig:orbits}
\end{figure}

\begin{figure*}
   \centering
\includegraphics[width=0.49\linewidth, trim=0mm 0mm 3mm 0mm, angle=0]{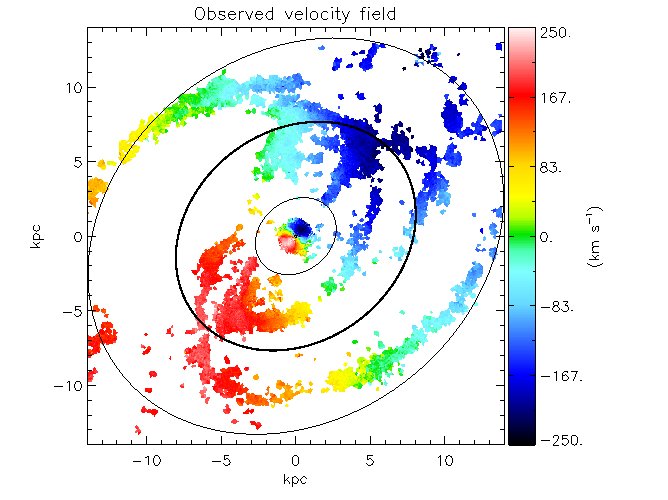}%\\ \vspace{0.4cm}
\includegraphics[width=0.49\linewidth, trim=0mm 0mm 3mm 0mm, angle=0]{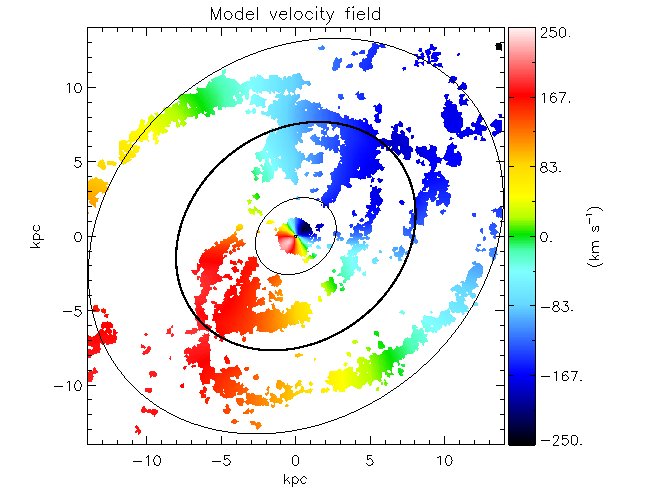}\\ \vspace{0.4cm}
\includegraphics[width=0.49\linewidth, trim=0mm 0mm 3mm 0mm, angle=0]{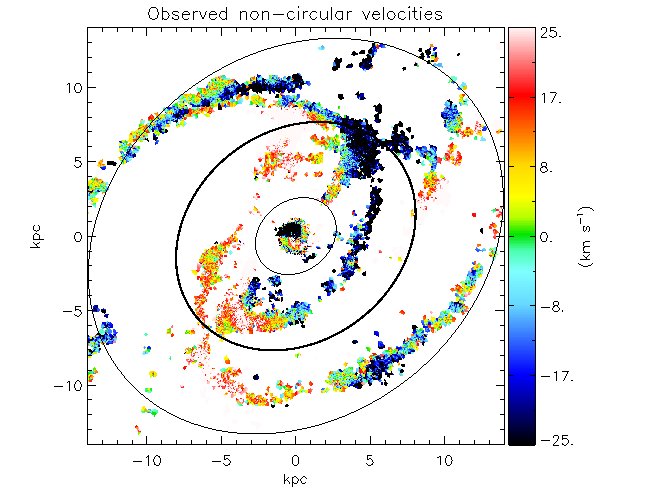}%\\ \vspace{0.4cm}
\includegraphics[width=0.49\linewidth, trim=0mm 0mm 3mm 0mm, angle=0]{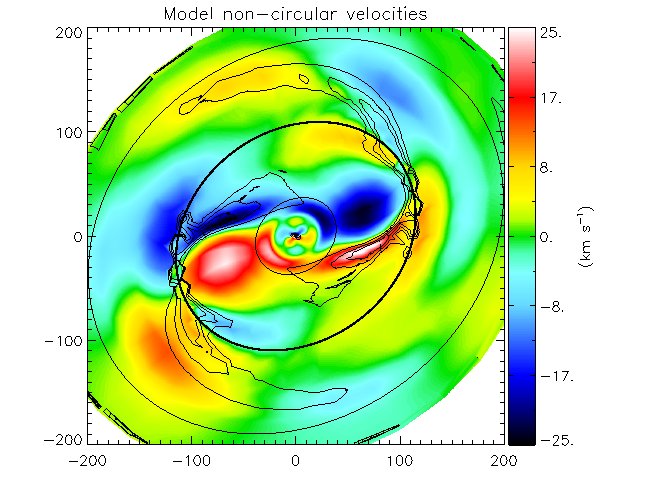}\\ \vspace{0.4cm}
\includegraphics[width=0.49\linewidth, trim=0mm 0mm 3mm 0mm, angle=0]{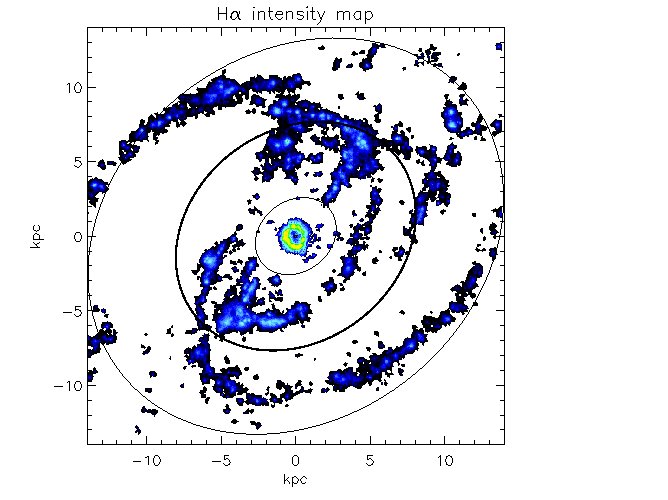}%\\ \vspace{0.4cm}
\includegraphics[width=0.49\linewidth, trim=0mm 0mm 3mm 0mm, angle=0]{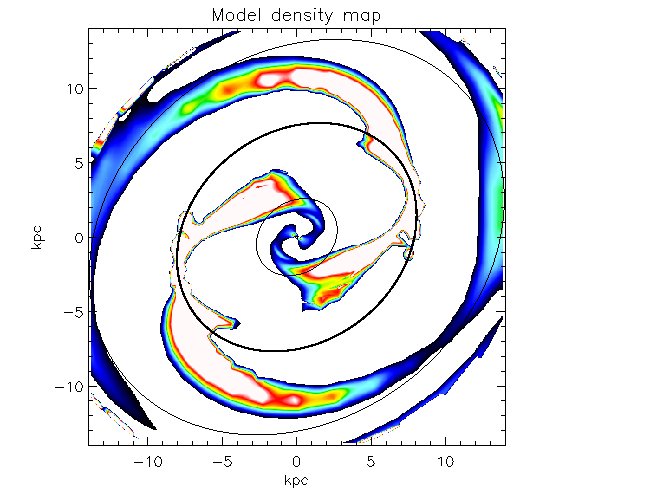}%\\ \vspace{0.4cm}
\caption{Black ellipses mark the resonance radii and the corotation (iILR at 90~pc, oILR at 2.9~kpc, CR at 8.6~kpc and OLR at 14.9~kpc). \emph{Upper left:} Observed velocity field observed . \emph{Upper right:} Velocity field generated by \Epic\ for a pattern speed of 36~\kmskpc. \emph{Middle left:} Observed non-circular velocities. \emph{Middle right:} Model non-circular velocities. \emph{Lower left:} Observed H$\alpha$ intensity map. \emph{Lower right:} Surface density map generated by \Epic, where the ratio between the perturbed and unperturbed surface density is illustrated. The model non-circular velocity map clearly displays the expected three-fold symmetric non-circular motions due to the presence of an $m=2$ gravitational perturbation outside corotation, and a pair of receding and approaching arms close to the oILR.}
\label{fig:fantomresult}
\end{figure*}

\begin{figure*}
   \centering
\includegraphics[width=0.52\linewidth, trim=10mm 0mm 5mm 0mm, angle=0]{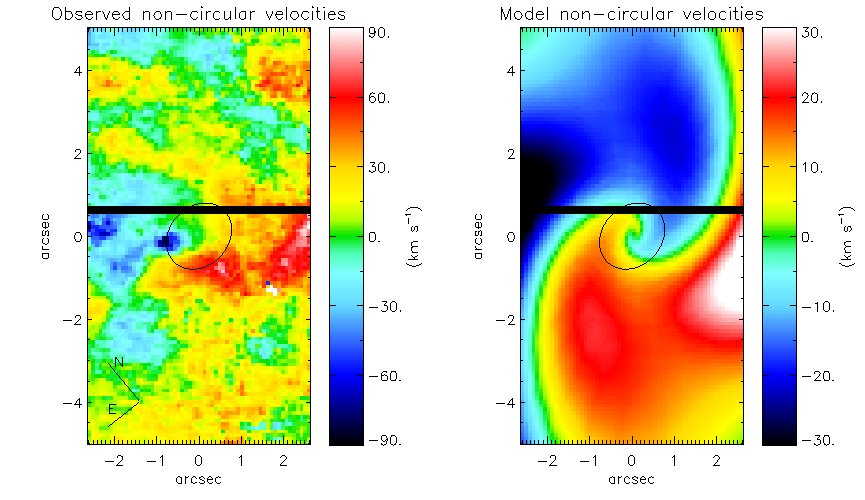}%\\ \vspace{0.4cm}
\includegraphics[width=0.46\linewidth, trim=0mm 0mm 0mm 10mm, angle=0]{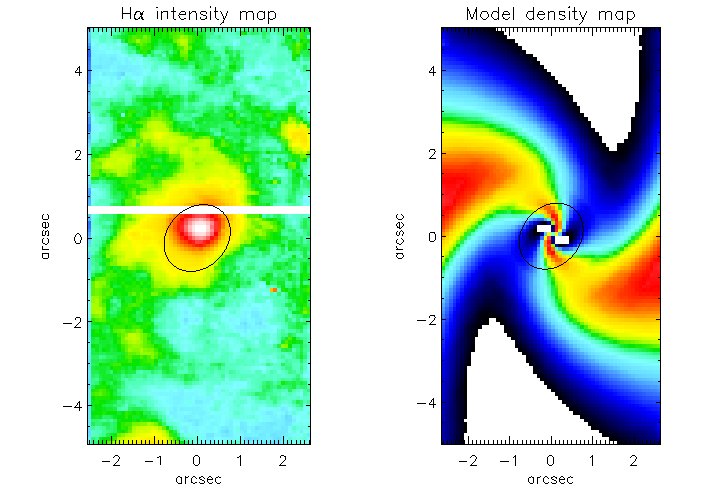}%\\ \vspace{0.4cm}
\caption{Black ellipse marks the resonance radii at the inner inner Lindblad resonance (60~pc) for a pattern speed of 36~\kmskpc. \emph{Left:} Observed non-circular velocities and model non-circular velocities. \emph{Right:} H$\alpha$ intensity map observed with GMOS and surface density map generated by \Epic, where the ratio between the perturbed and unperturbed surface density is illustrated.}
\label{fig:gmosresult}
\end{figure*}

\section{The pattern speed of NGC\,1097}
\begin{figure}
   \centering
\includegraphics[width=0.75\linewidth, trim=0mm 0mm 0mm 0mm, angle=-90]{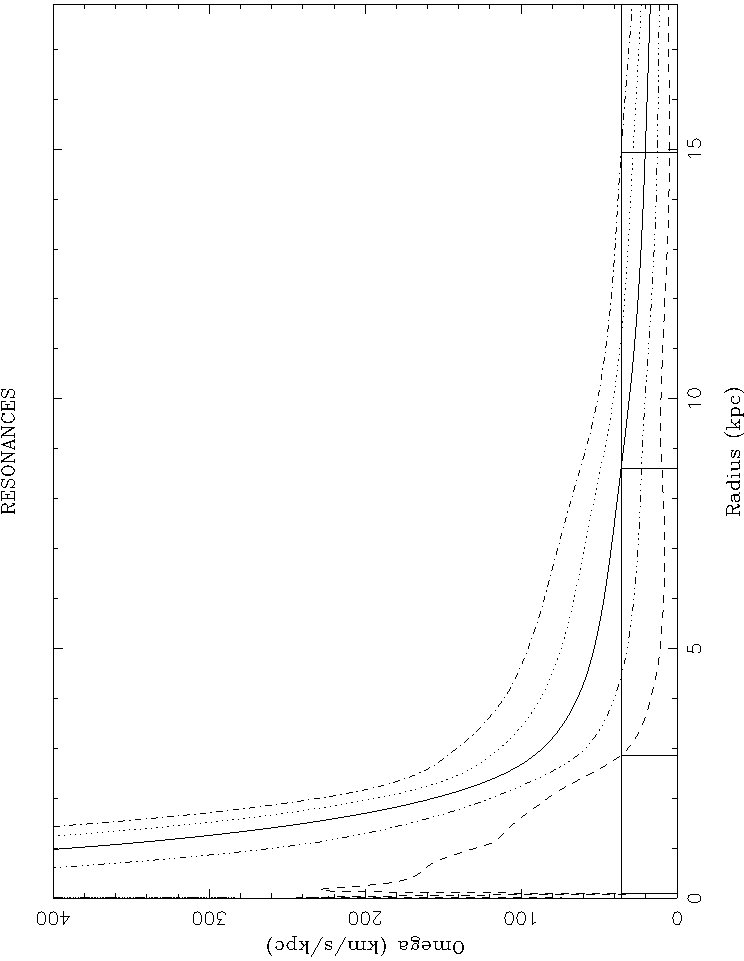}%\\ \vspace{0.4cm}
\caption{Angular velocity $\Omega$, against radius derived from the rotation curve (solid line). The dashed line shows $\Omega- \kappa/2$, the dotted-dashed line shows $\Omega+ \kappa/2$, the dotted and dashed-triple dotted lines show  $\Omega\pm \kappa/4$. The pattern speed of  36~km s$^{-1}$ kpc$^{-1}$ is shown by the horizontal solid line, and the resonances are indicated by the vertical lines.}
\label{fig:angfreq}
\end{figure}

The estimation of the pattern speed of NGC\,1097 using \Epic\ gives a value of 36$\pm2$~\kmskpc. This frequency places corotation at 8.6$\pm$0.5~kpc, two inner Lindblad resonances at 60$\pm$~5~pc and 2.9$\pm$0.1~kpc and the outer Lindblad resonance at 14.9$\pm$0.9~kpc (see Fig.~\ref{fig:angfreq}). These resonance locations are obtained assuming a linear approximation, therefore they may slightly change in the presence of a strong bar. Their location is marked in Figs.~\ref{fig:fantomresult} by black ellipses, where one can observe that the corotation is placed around the end of the bar and that the nuclear star forming ring, with radius around 1~kpc, is located between the two inner Lindblad resonances. 

The errors of the resonance positions are functions of the errors of the rotation curve and its derivatives. However, we have estimated these errors taking into account the error in the pattern speed, which gives an uncertainty at least one order of magnitude larger. Still the error of the pattern speed may be underestimated if systematic errors are involved, what would also increase the errors on the resonance positions. Another error source that we have to mention is the uncertainty in the rotational curve between $\sim$1.5--3~kpc due to the lack of pixels covering the region of interest.

The pattern speed value derived using \Epic\ is in full agreement with the value used in \citet{vandeVenFathi2010} and with the value obtained by using the model-independent Tremaine Weinberg method \citep{Tremaine1984}. The latter is the result of the Tremaine Weinberg method applied to the \fantomm\ velocity field weighted with the Spitzer image \citep{Fathi2009}, i.e. $30\pm8$~\kmskpc. 

Alternatively, we use the phase-reversal method \citep{Font2011} in order to determine the corotation radius and the associated bar pattern speed. In short, the method locates the $\pi$ phase reversals of the non-circular velocities. This way, we obtain the distribution of the phase reversals as a function of the galactocentric radius. The radial distribution shows several peaks, which can be fitted with standard Gaussian functions. Assuming that streaming velocities change sign at the corotation radius, we can identify the maximum in the radial distribution histogram with the corotation radius. Applying the phase reversal method to the residual velocity map of NGC\,1097, the strongest peak of the phase reversal histogram is found at a radius of $7.7\pm0.2$ kpc, and the corresponding pattern speed is $39.5\pm1.1$~\kmskpc. It should be noted the errors in the latter value only include formal errors and that they are expected to be larger when including the systamic uncertainties involved. Nevertheless, this is a good match to the value of the pattern speed derived by means of the analytic dynamical model (see section~\ref{sec:3}).

In order to compare the estimated pattern speed of this galaxy with simulations, we estimate the bar length, $a$, of NGC\,1097. We study this parameter by looking at the ellipticity of the galaxy, at the phase shift of the second Fourier component and at the contrast parameter defined by \citet{1990ApJ...357...71O}. We find that $a$ is approximately $7.9\pm0.6$~kpc. This gives a ratio of the radius of corotation, CR-over-bar length, $\mathcal{R} = R_{\rm CR}/a $, of 1.0--1.2, which corresponds to a fast bar. From numerical models, it is known that in barred galaxies $\mathcal{R}$ has a values of $1.2\pm2$ \citep[see e.g.][]{Athanassoula1992b}, which is in concordance with our estimations for NGC\,1097.

On the other hand, since it is widely believed that nuclear rings form at inner Lindbland resonances \citep[e.g.][]{1996FCPh...17...95B}, a faster pattern speed could have been expected. However, based on both the ionized gas and the neutral gas rotation curve, we find that a pattern speed that would locate the nuclear ring at the outer ILR, will locate the corotation radius  in an unrealistically smaller radius in the galaxy and consequently produce $\mathcal{R}\ll0.9$ ($\mathcal{R}\sim0.3$).

\section{Comparison of observations and the model}

To compare our high-resolution Fabry-Perot kinematic maps with the modelled velocity fields, we first look at the non-circular velocity maps. Figures~\ref{fig:fantomresult}~and~\ref{fig:gmosresult} show the observed velocity field, non-circular velocities and H$\alpha$ luminosities of NGC\,1097 together with the analytically modelled velocities and densities. At large scales, our model follows the outer spiral arms of the galaxy until the end, almost closing after the OLR. Furthermore, both the observed and model non-circular velocity fields reveal three kinematic spiral arms after the corotation radius, at 8.6$\pm$0.5~kpc. These arms are produced by the bi-symmetric component of the gravitational potential. Inside corotation, our model is able to reproduce the expected velocity jump linking the end of the bar with the outer galactic spiral arms. This is consistent with similar velocity jumps found by \citet{2006MNRAS.369L..56P}. Moreover, the correlation between model and observations here becomes less evident than after corotation since our model presents the dust lane rotated with respect the central lane of NGC\,1097. This may be due to the fact that \Epic\ reproduces a weak bar model while the bar in NGC\,1097 would easily qualify as a strong bar, and it is expected that dust lanes are more straight as they get stronger \citep{Athanassoula1992b,2002MNRAS.337..808K,2009ApJ...706L.256C}. Also in the bar region, although the modelled dust lanes are rotated with respect the observed central ones, we find signatures of strong shocks along the front edge of the bar, where the gas crossing the bar sharply bends inward along the dust lanes towards the centre. They are very important since they bend the orbits across the bar sharply inwards and thus help to feed the nuclear region.

We further look into the central region by comparing the analytic model with the GMOS-IFU velocity field. Our model reproduces the density structures corresponding to the two strong nuclear spiral arms found by \citet{vandeVenFathi2010}. In the density panels of Figs.~\ref{fig:fantomresult}~and~\ref{fig:gmosresult}, the modelled surface density is plotted as the ratio between the perturbed and unperturbed surface density. In the inner tens of parsecs, two spiral arms are modelled. These arms are not obvious in the ionized gas maps, predominantly due to strong contamination from the diffuse ionized gas emission. Nevertheless, two clear nuclear spiral arms are seen in wavelet analysis maps presented by \citet[][their Fig.~1]{Lou2001} and the structure map shown by \cite[][middle panel of their Figure 3]{Fathi2006}. We further note that \citet[][]{Prieto2005} argued for the presence of three spiral arms at the centre of this galaxy. 
However, quantitative kinematic analysis highlights the presence of three kinematic spiral arms (i.e., two arms in morphology), in \citet{vandeVenFathi2010}. In our kinematic analysis shown in Fig.~\ref{fig:gmosresult}, using the same dataset, these three kinematic spiral arms are not as clear as in their map due to the slightly different rotational curve used. Moreover, our model predicts one kinematic spiral arm, as shown in the second panel of Fig.~\ref{fig:gmosresult}. That difference between the observations and the model is produced by the fact that in the observations, the  shape of the non-circular velocities depend on the derived rotation curve that we are subtracting. That is because in these circular velocities part of the $m=$1 component is hidden and this is the main reason why it is observed 3 kinematic spiral arms instead of 1 as theory indicates. In our modelled non-circular velocity map, we observe one arm in the absence of hidden $m=$1 contribution in the rotation curve. In addition, looking at our modelled non-circular velocity map, we can observe a $m=$3 contribution hidden in the map that is seen due to the uneven continuity in the beginning of the one spiral. 
This predominance of the $m=$3 and $m=$1 Fourier term of the sight velocity in the non-circular velocities, supports the idea of a situation in which the gravitational potential is bi-symmetrically perturbed (see first paragraph of section 3.1 above here). %In \mbox{Fig.~\ref{fig:fantomresult}-\emph{right middle panel}} and Fig.~\ref{fig:gmosresult}, we also note the presence of strong shocks along the front side of the nuclear spiral arms, where the gas crossing the arms (rotating faster than the nuclear spirals pattern speed) sharply bends inward along the lanes towards the centre.

Observations also reveal that the nuclear spiral arms start at the nuclear ring located at around 1~kpc. In our model, these arms also start at the position of the nuclear ring, which is a structure that is the prolongation of the dust lanes. Furthermore, the ring is not entirely closed in our model and it is not located anywhere near the location of any main resonances.

\subsection{Is the ring migrating inwards?}

Our analysis suggests that what is commonly referred to as the circumnuclear ring in NGC\,1097 %is not a ring in the strict dynamical sense, but that it 
is a region of enhanced star formation activity at the inner ends of the dust lanes that cut through the major axis of the primary bar, crossing the outer ILR and connecting with the nuclear spiral arms. %In any case, what is completely clear is that what

The understanding of the exact location of the ring is important in order to comprehend the effects of secular evolution of NGC\,1097. Circumnuclear rings are commonly thought to be located at or close to the ILRs \citep[e.g.][]{1994AJ....108..476B,Elmegreen1994,1996FCPh...17...95B}. However, considering the locations of the inner and outer ILRs in our epicyclic model, this ring seems not to be located near either, but in a region significantly distant from the two. The resonance radii that we have described are only valid under a weak non-axisymmetric perturbation, while NGC\,1097 has a strong bar. However, this estimation for the resonance radii are likely to be a useful indication \citep{1995ApJ...454..623K,2011ApJ...739..104M}. This is the same situation found in other galaxies like NGC\,1068 \citep{1988ApJ...334..573T}, M\,100 \citep{1988A&A...200...29A} or NGC\,3504 \citep{1993ApJ...418..687K}. 

Another preferred general location for circumnuclear rings is the turnover location in the rotation curve, where the rotation curve goes from a initial steeply raising to a flatter segment \citep[e.g.][and references therein]{1996FCPh...17...95B}. The circumnuclear star forming ring of NGC\,1097 is located at $\sim$1~kpc which approximately coincides with the turnover radius \citep[see][and \mbox{Fig.~\ref{fig:rot}-\emph{right top panel}}]{1979NZJS...22..325W}. Hence, our results agree with this situation but place the circumnuclear ring between the two ILRs. This may point out an inward migration of the ring from the oILR.

This scenario has been predicted by \citet{2000ApJ...529..109F}, \citet{2003ApJ...582..723R} and \citet{vandeVenChang2009}, who have shown evidence, by simulations or analytical studies, that rings could form at a Lindblad resonance, but that they may migrate inward through the galactic disc. These studies find that the migration of the circumnuclear ring is produced by the inflowing gas from the dust lane to the ring, by the underlined gravitational potential and by the viscosity. However, tests made by \citet{2003ApJ...582..723R} in their simulations seem to indicate that the main responsible mechanism is the gas flowing down from the dust lanes. 
Notably, the latter study predicts that when the ring is not located at a Lindblad resonance, inward migration may well lead to strong star formation due to gas accumulation at its edges producing enhancements in the density and favouring unstable scenarios against self-gravitation. Hence, as the ring moves, the star formation along the ring changes from probably presenting an aged azimuthal gradient, produced by a location of the star formation at the connection points of the dust lanes with the ring, to a randomly distributed star formation \citep{vandeVenChang2009}. This scenario is also confirmed by the randomly distributed and strong star formation found in the circumnuclear ring of NGC\,1097 with the no-clear age azimuthal gradient \citep{Sandstrom2010,2011ApJ...736..129H,pinol2011}. %\citet{vandeVenChang2009} also predicts a no clear age azimuthal gradient along the ring produced by an increase of the gas density at the edges of the ring produced as it goes inward 
 
%This scenario leaves open the problem of how such enhanced and ordered star formation could occur in such a, dynamically speaking, unremarkable location. One possible solution for this is can de ven \& Chung... {\bf do the predict if the ring dissolves when it migrates? If their study offers a solution, please add at least a paragraph about their study.}

\section{Conclusions}
\label{sec:conclusions}

We present here an unprecedented galactic-scale dynamical model of the kinematics and morphology of NGC\,1097. Our analytic model has been constructed by deriving the gravitational potential of NGC\,1097 from the old stellar population and observed rotation curve, which we use to generate the solution of the equations of motion of the gas in this galaxy, assuming the epicyclic approximation. We verify the model by applying it to high-resolution (0\farcs83/pix) Fabry-Perot interferometric data covering the entire galaxy. We further zoom into the central few hundred parsecs by using high spatial resolution (0\farcs1/pix) two-dimensional spectroscopy of the nuclear central kiloparsec. The calculation of the solution of the equations of motion is done by using our own customized code \Epic\footnote{The \Epic\ code can be obtained from npi@astro.su.se . }.

\begin{table}
 \caption{A summary of the physical parameters derived in this work.}
   \label{tab:1097mod}
    \centering   
      \begin{threeparttable}
%\scalebox{4}{
         \begin{tabularx}{\linewidth}{|X|X|}
%         \toprule
	\hline
	\multicolumn{2}{c}{NGC\,1097} \\
	\hline
%             \midrule
	 \gray $V_{\rm sys}$ & $1283\pm10$~\kms\\
	P.A. & $128\pm 10^{\circ}$\\
	 \gray Bar length & $7.9\pm0.6$~kpc \\		
	Pattern speed  & \multirow{2}{*}{$39.5\pm1.1$~\kmskpc} \\ 
	(Phase reversal method)$^a$ &\\
	\gray  Pattern speed (\Epic) & $36\pm2$~\kmskpc \\	
	Pattern speed  & \multirow{2}{*}{$30\pm8$~\kmskpc}\\
	(TW method)$^{b}$&\\
	\gray CR & $8.6\pm0.5$~kpc\\
	IILR and OILR & $60\pm5$~pc and $2.9\pm0.1$~kpc\\
	\gray OLR & $14.9\pm0.9$~kpc\\
%         \bottomrule
	\hline
       \end{tabularx}%}
    \begin{tablenotes}
    \item[a] \citet{Font2011}, \item[b] \citet{Tremaine1984,Fathi2009}
%$^a$ \citet{2004A&A...415..941E}, $^b$ \citet{1988ngc..book.....T}
    \end{tablenotes}
 \end{threeparttable}
\end{table}

We find that the bar in NGC\,1097 has a radius of 7.9$\pm$0.6~kpc and that it is rotating with a pattern speed of 36$\pm2$~\kmskpc. This pattern speed places the corotation radius at 8.6$\pm$0.5~kpc, the outer Lindblad resonance at 14.9$\pm$0.9~kpc and two inner Lindblad resonances at 60$\pm$5~pc and 2.9$\pm$0.1~kpc, based on the epicyclic approximation. These derivations lead to a ratio of CR-over-bar length, $\mathcal{R}$, of 1.0--1.2, which is in concordance with numerical simulations and models of barred galaxies \citep[e.g.][]{Athanassoula1992b}. Our analytic dynamical model reproduces all the significant kinematic and structural signatures:

\begin{enumerate}
  \setcounter{enumi}{0}
\item The outer two spiral arms that are created at corotation and almost close after the outer Lindblad resonance. Also the related three-fold symmetric non-circular motions are recovered in our model.
\item The expected velocity jump linking the end of the bar with the outer galactic spiral arms, consistent with similar velocity jumps found by \citet{2006MNRAS.369L..56P}.
\item The kinematics produced by two nuclear spiral arms seen in wavelet analysis maps presented by \citet[][their Fig.~1]{Lou2001}, the structure map shown in \cite{Fathi2006} and the kinematics study presented in \citet{vandeVenFathi2010}.
\item The position of the starting point of the nuclear spiral arms.
\item The location of the circumnuclear ring.
\end{enumerate}

Our model also reveals that the circumnuclear ring of this galaxy is not located near any of the inner Lindblad resonance radii, as commonly expected for nuclear rings in barred galaxies. We find compelling evidence that the ring is located in the region between the inner and the outer ILRs. This may indicate that the ring once formed at the outer inner Lindblad resonance radius, and it has been migrating toward the centre of the galactic gravitational potential. The plausibility of such ring migration has been previously predicted by \citet{2000ApJ...529..109F}, \citet{2003ApJ...582..723R} and \citet{vandeVenChang2009}.

\section*{ACKNOWLEDGMENTS}
We wish to thank to Per-Olof Lindblad and Panos Patsis for their insightful and invaluable comments and James Higdon for kindly providing his \Hi\ data. NP-F is supported by the Nordic Optical Telescope Scientific Association (NOTSA). KF acknowledges support from the Swedish Research Council (Vetenskapsr\aa det) and the Swedish Royal Academy of Sciences' Crafoord Foundation. KF also acknowledges the hospitality of the Max-Planck-Institut f\"ur extraterrestrische Physik (MPE) where parts of this work were carried out.

%\begin{thebibliography}{}
%\bibitem[\protect\citeauthoryear{Fathi et al.}{2006}]{f06} Fathi, K. et al. 2006, ApJ, 641, L25
%\bibitem[\protect\citeauthoryear{Hsieh et al.}{2012}]{hsieh12}  Hsieh, P-Y et al. 2012, ApJ, 747, 90
%\bibitem[\protect\citeauthoryear{Hsieh et al.}{2011}]{hsieh11}  Hsieh, P-Y et al. 2011, ApJ, 736, 129
%\bibitem[\protect\citeauthoryear{Ondrechen et al.}{1989}]{o89} Ondrechen, M. P., van der Hulst, J. M., Hummel, E. 1989, ApJ, 342, 39
%\bibitem[\protect\citeauthoryear{Pi\~no-Ferrer et al.}{2013}]{npf13} Pi\~nol-Ferrer, N., Fathi, K., Hernandez, O., Carignan, C. 2013, in preparation 
%\bibitem[\protect\citeauthoryear{Pi\~no-Ferrer et al.}{2011}]{npf11} Pi\~nol-Ferrer, N., Fathi, K., Lundgren, A. A., van de Ven, G. 2011, MNRAS, 414, 529 
%\bibitem[\protect\citeauthoryear{Sersic}{1958}]{sersic58} Sersic, J. L. 1958, The Observatory, 78, 123 
%\bibitem[\protect\citeauthoryear{Storchi-Bergmann et al.}{1993}]{sb93} Storchi-Bergmann, T., Baldwin, J. A., Wilson, A. S. 1993, ApJ, 410, L11
%\bibitem[\protect\citeauthoryear{Storchi-Bergmann et al.}{2003}]{sb03} Storchi-Bergmann, T. et al. 2003, ApJ, 598, 956
%\bibitem[\protect\citeauthoryear{van der Kruit \& Freeman}{2011}]{vanderkruit11} van der Krui P.C. \& Freeman K. 2011, \araa, 49, 301

%\end{thebibliography}

\bibliographystyle{mn2e}
\bibliography{mybib}

\begin{thebibliography}{}

\bibitem[\protect\citeauthoryear{{Arsenault}, {Boulesteix}, {Georgelin} \&
  {Roy}}{{Arsenault} et~al.}{1988}]{1988A&A...200...29A}
{Arsenault} R.,  {Boulesteix} J.,  {Georgelin} Y.,    {Roy} J.-R.,  1988, \aap,
  200, 29

\bibitem[\protect\citeauthoryear{{Athanassoula}}{{Athanassoula}}{1992}]{Athana%
ssoula1992b}
{Athanassoula} E.,  1992, \mnras, 259, 345

\bibitem[\protect\citeauthoryear{{Baker}}{{Baker}}{2000}]{2000PhDT.........6B}
{Baker} A.~J.,  2000

\bibitem[\protect\citeauthoryear{{Binney} \& {Tremaine}}{{Binney} \&
  {Tremaine}}{2008}]{BinneyTremaine2008}
{Binney} J.,  {Tremaine} S.,  2008, {Galactic Dynamics: Second Edition}.
Princeton University Press

\bibitem[\protect\citeauthoryear{{Boone}, {Baker}, {Schinnerer}, {Combes},
  {Garc{\'{\i}}a-Burillo}, {Neri}, {Hunt}, {L{\'e}on}, {Krips}, {Tacconi} \&
  {Eckart}}{{Boone} et~al.}{2007}]{2007A&A...471..113B}
{Boone} F.,  {Baker} A.~J.,  {Schinnerer} E.,  {Combes} F.,
  {Garc{\'{\i}}a-Burillo} S.,  {Neri} R.,  {Hunt} L.~K.,  {L{\'e}on} S.,
  {Krips} M.,  {Tacconi} L.~J.,    {Eckart} A.,  2007, \aap, 471, 113

\bibitem[\protect\citeauthoryear{{Bournaud}, {Combes}, {Jog} \&
  {Puerari}}{{Bournaud} et~al.}{2005}]{2005A&A...438..507B}
{Bournaud} F.,  {Combes} F.,  {Jog} C.~J.,    {Puerari} I.,  2005, \aap, 438,
  507

\bibitem[\protect\citeauthoryear{{Buta} \& {Combes}}{{Buta} \&
  {Combes}}{1996}]{1996FCPh...17...95B}
{Buta} R.,  {Combes} F.,  1996, \fcp, 17, 95

\bibitem[\protect\citeauthoryear{{Byrd}, {Rautiainen}, {Salo}, {Buta} \&
  {Crocher}}{{Byrd} et~al.}{1994}]{1994AJ....108..476B}
{Byrd} G.,  {Rautiainen} P.,  {Salo} H.,  {Buta} R.,    {Crocher} D.~A.,  1994,
  \aj, 108, 476

\bibitem[\protect\citeauthoryear{{Canzian}}{{Canzian}}{1993}]{Canzian1993}
{Canzian} B.,  1993, \apj, 414, 487

\bibitem[\protect\citeauthoryear{{Combes}}{{Combes}}{2001}]{2001sac..conf..223%
C}
{Combes} F.,  2001, p.~223

\bibitem[\protect\citeauthoryear{{Comer{\'o}n}, {Mart{\'{\i}}nez-Valpuesta},
  {Knapen} \& {Beckman}}{{Comer{\'o}n} et~al.}{2009}]{2009ApJ...706L.256C}
{Comer{\'o}n} S.,  {Mart{\'{\i}}nez-Valpuesta} I.,  {Knapen} J.~H.,
  {Beckman} J.~E.,  2009, \apjl, 706, L256

\bibitem[\protect\citeauthoryear{{Davies}, {Maciejewski}, {Hicks}, {Tacconi},
  {Genzel} \& {Engel}}{{Davies} et~al.}{2009}]{Davies2009}
{Davies} R.~I.,  {Maciejewski} W.,  {Hicks} E.~K.~S.,  {Tacconi} L.~J.,
  {Genzel} R.,    {Engel} H.,  2009, \apj, 702, 114

\bibitem[\protect\citeauthoryear{{Davies}, {M{\"u}ller S{\'a}nchez}, {Genzel},
  {Tacconi}, {Hicks}, {Friedrich} \& {Sternberg}}{{Davies}
  et~al.}{2007}]{2007ApJ...671.1388D}
{Davies} R.~I.,  {M{\"u}ller S{\'a}nchez} F.,  {Genzel} R.,  {Tacconi} L.~J.,
  {Hicks} E.~K.~S.,  {Friedrich} S.,    {Sternberg} A.,  2007, \apj, 671, 1388

\bibitem[\protect\citeauthoryear{{Dicaire}, {Carignan}, {Amram}, {Hernandez},
  {Chemin}, {Daigle}, {de Denus-Baillargeon}, {Balkowski}, {Boselli}, {Fathi}
  \& {Kennicutt}}{{Dicaire} et~al.}{2008}]{Dicaire2008}
{Dicaire} I.,  {Carignan} C.,  {Amram} P.,  {Hernandez} O.,  {Chemin} L.,
  {Daigle} O.,  {de Denus-Baillargeon} M.-M.,  {Balkowski} C.,  {Boselli} A.,
  {Fathi} K.,    {Kennicutt} R.~C.,  2008, \mnras, 385, 553

\bibitem[\protect\citeauthoryear{{Elmegreen}}{{Elmegreen}}{1994}]{Elmegreen199%
4}
{Elmegreen} B.~G.,  1994, \apjl, 425, L73

\bibitem[\protect\citeauthoryear{{Erwin}}{{Erwin}}{2004}]{2004A&A...415..941E}
{Erwin} P.,  2004, \aap, 415, 941

\bibitem[\protect\citeauthoryear{{Fathi}, {Beckman}, {Pi{\~n}ol-Ferrer},
  {Hernandez}, {Mart{\'{\i}}nez-Valpuesta} \& {Carignan}}{{Fathi}
  et~al.}{2009}]{Fathi2009}
{Fathi} K.,  {Beckman} J.~E.,  {Pi{\~n}ol-Ferrer} N.,  {Hernandez} O.,
  {Mart{\'{\i}}nez-Valpuesta} I.,    {Carignan} C.,  2009, \apj, 704, 1657

\bibitem[\protect\citeauthoryear{{Fathi}, {Storchi-Bergmann}, {Riffel},
  {Winge}, {Axon}, {Robinson}, {Capetti} \& {Marconi}}{{Fathi}
  et~al.}{2006}]{Fathi2006}
{Fathi} K.,  {Storchi-Bergmann} T.,  {Riffel} R.~A.,  {Winge} C.,  {Axon}
  D.~J.,  {Robinson} A.,  {Capetti} A.,    {Marconi} A.,  2006, \apjl, 641, L25

\bibitem[\protect\citeauthoryear{{Fathi}, {van de Ven}, {Peletier}, {Emsellem},
  {Falc{\'o}n-Barroso}, {Cappellari} \& {de Zeeuw}}{{Fathi}
  et~al.}{2005}]{Fathi2005}
{Fathi} K.,  {van de Ven} G.,  {Peletier} R.~F.,  {Emsellem} E.,
  {Falc{\'o}n-Barroso} J.,  {Cappellari} M.,    {de Zeeuw} T.,  2005, \mnras,
  364, 773

\bibitem[\protect\citeauthoryear{{Font}, {Beckman}, {Epinat}, {Fathi},
  {Guti{\'e}rrez} \& {Hernandez}}{{Font} et~al.}{2011}]{Font2011}
{Font} J.,  {Beckman} J.~E.,  {Epinat} B.,  {Fathi} K.,  {Guti{\'e}rrez} L.,
  {Hernandez} O.,  2011, \apjl, 741, L14

\bibitem[\protect\citeauthoryear{{Fridman} \& {Khoruzhii}}{{Fridman} \&
  {Khoruzhii}}{2003}]{2003SSRv..105....1F}
{Fridman} A.~M.,  {Khoruzhii} O.~V.,  2003, \ssr, 105, 1

\bibitem[\protect\citeauthoryear{{Fukuda}, {Habe} \& {Wada}}{{Fukuda}
  et~al.}{2000}]{2000ApJ...529..109F}
{Fukuda} H.,  {Habe} A.,    {Wada} K.,  2000, \apj, 529, 109

\bibitem[\protect\citeauthoryear{{Garc{\'{\i}}a-Burillo}, {Combes},
  {Schinnerer}, {Boone} \& {Hunt}}{{Garc{\'{\i}}a-Burillo}
  et~al.}{2005}]{2005A&A...441.1011G}
{Garc{\'{\i}}a-Burillo} S.,  {Combes} F.,  {Schinnerer} E.,  {Boone} F.,
  {Hunt} L.~K.,  2005, \aap, 441, 1011

\bibitem[\protect\citeauthoryear{{Hernandez}, {Gach}, {Carignan} \&
  {Boulesteix}}{{Hernandez} et~al.}{2003}]{Hernandez2003}
{Hernandez} O.,  {Gach} J.-L.,  {Carignan} C.,    {Boulesteix} J.,  2003, in
  {M.~Iye \& A.~F.~M.~Moorwood} ed., Society of Photo-Optical Instrumentation
  Engineers (SPIE) Conference Series Vol.~4841 of Society of Photo-Optical
  Instrumentation Engineers (SPIE) Conference Series, {FaNTOmM: Fabry Perot of
  New Technology for the Observatoire du mont Megantic}.
pp 1472--1479

\bibitem[\protect\citeauthoryear{{Hicks}, {Davies}, {Malkan}, {Genzel},
  {Tacconi}, {M{\"u}ller S{\'a}nchez} \& {Sternberg}}{{Hicks}
  et~al.}{2009}]{2009ApJ...696..448H}
{Hicks} E.~K.~S.,  {Davies} R.~I.,  {Malkan} M.~A.,  {Genzel} R.,  {Tacconi}
  L.~J.,  {M{\"u}ller S{\'a}nchez} F.,    {Sternberg} A.,  2009, \apj, 696, 448

\bibitem[\protect\citeauthoryear{{Higdon} \& {Wallin}}{{Higdon} \&
  {Wallin}}{2003}]{Higdon2003}
{Higdon} J.~L.,  {Wallin} J.~F.,  2003, \apj, 585, 281

\bibitem[\protect\citeauthoryear{{Hsieh}, {Matsushita}, {Liu}, {Ho}, {Oi} \&
  {Wu}}{{Hsieh} et~al.}{2011}]{2011ApJ...736..129H}
{Hsieh} P.-Y.,  {Matsushita} S.,  {Liu} G.,  {Ho} P.~T.~P.,  {Oi} N.,    {Wu}
  Y.-L.,  2011, \apj, 736, 129

\bibitem[\protect\citeauthoryear{{Hummel}, {van der Hulst}, {Kennicutt} \&
  {Keel}}{{Hummel} et~al.}{1990}]{1990A&A...236..333H}
{Hummel} E.,  {van der Hulst} J.~M.,  {Kennicutt} R.~C.,    {Keel} W.~C.,
  1990, \aap, 236, 333

\bibitem[\protect\citeauthoryear{{Kenney}, {Carlstrom} \& {Young}}{{Kenney}
  et~al.}{1993}]{1993ApJ...418..687K}
{Kenney} J.~D.~P.,  {Carlstrom} J.~E.,    {Young} J.~S.,  1993, \apj, 418, 687

\bibitem[\protect\citeauthoryear{{Kennicutt} et~al.,}{{Kennicutt}
  et~al.}{2003}]{Kennicutt2003}
{Kennicutt} J.,  et~al., 2003, \pasp, 115, 928

\bibitem[\protect\citeauthoryear{{Knapen}, {Beckman}, {Heller}, {Shlosman} \&
  {de Jong}}{{Knapen} et~al.}{1995}]{1995ApJ...454..623K}
{Knapen} J.~H.,  {Beckman} J.~E.,  {Heller} C.~H.,  {Shlosman} I.,    {de Jong}
  R.~S.,  1995, \apj, 454, 623

\bibitem[\protect\citeauthoryear{{Knapen}, {P{\'e}rez-Ram{\'{\i}}rez} \&
  {Laine}}{{Knapen} et~al.}{2002}]{2002MNRAS.337..808K}
{Knapen} J.~H.,  {P{\'e}rez-Ram{\'{\i}}rez} D.,    {Laine} S.,  2002, \mnras,
  337, 808

\bibitem[\protect\citeauthoryear{{Lindblad}, {Lindblad} \&
  {Athanassoula}}{{Lindblad} et~al.}{1996}]{Lindblad1996}
{Lindblad} P.~A.~B.,  {Lindblad} P.~O.,    {Athanassoula} E.,  1996, \aap, 313,
  65 (LLA96)

\bibitem[\protect\citeauthoryear{{Lindblad} \& {Lindblad}}{{Lindblad} \&
  {Lindblad}}{1994}]{Lindblad1994}
{Lindblad} P.~O.,  {Lindblad} P.~A.~B.,  1994, in {I.~R.~King} ed., Physics of
  the Gaseous and Stellar Disks of the Galaxy, ASP Conference Series Vol.~66,
  {}.
p.~29

\bibitem[\protect\citeauthoryear{{Lou}, {Yuan}, {Fan} \& {Leon}}{{Lou}
  et~al.}{2001}]{Lou2001}
{Lou} Y.-Q.,  {Yuan} C.,  {Fan} Z.,    {Leon} S.,  2001, \apjl, 553, L35

\bibitem[\protect\citeauthoryear{{Martini}, {Regan}, {Mulchaey} \&
  {Pogge}}{{Martini} et~al.}{2003}]{Martini2003}
{Martini} P.,  {Regan} M.~W.,  {Mulchaey} J.~S.,    {Pogge} R.~W.,  2003, \apj,
  589, 774

\bibitem[\protect\citeauthoryear{{Mazzuca}, {Swaters}, {Knapen} \&
  {Veilleux}}{{Mazzuca} et~al.}{2011}]{2011ApJ...739..104M}
{Mazzuca} L.~M.,  {Swaters} R.~A.,  {Knapen} J.~H.,    {Veilleux} S.,  2011,
  \apj, 739, 104

\bibitem[\protect\citeauthoryear{{Ohta}, {Hamabe} \& {Wakamatsu}}{{Ohta}
  et~al.}{1990}]{1990ApJ...357...71O}
{Ohta} K.,  {Hamabe} M.,    {Wakamatsu} K.-I.,  1990, \apj, 357, 71

\bibitem[\protect\citeauthoryear{{Patsis}}{{Patsis}}{2006}]{2006MNRAS.369L..56%
P}
{Patsis} P.~A.,  2006, \mnras, 369, L56

\bibitem[\protect\citeauthoryear{{Pi{\~n}ol-Ferrer}, {Fathi}, {Lundgren} \&
  {van de Ven}}{{Pi{\~n}ol-Ferrer} et~al.}{2011}]{pinol2011}
{Pi{\~n}ol-Ferrer} N.,  {Fathi} K.,  {Lundgren} A.,    {van de Ven} G.,  2011,
  \mnras, 414, 529

\bibitem[\protect\citeauthoryear{{Pi{\~n}ol-Ferrer}, {Lindblad} \&
  {Fathi}}{{Pi{\~n}ol-Ferrer} et~al.}{2012}]{PinolFerrer2012}
{Pi{\~n}ol-Ferrer} N.,  {Lindblad} P.~O.,    {Fathi} K.,  2012, \mnras, 421,
  1089

\bibitem[\protect\citeauthoryear{{Prieto}, {Maciejewski} \&
  {Reunanen}}{{Prieto} et~al.}{2005}]{Prieto2005}
{Prieto} M.~A.,  {Maciejewski} W.,    {Reunanen} J.,  2005, \aj, 130, 1472

\bibitem[\protect\citeauthoryear{{Regan} \& {Teuben}}{{Regan} \&
  {Teuben}}{2003}]{2003ApJ...582..723R}
{Regan} M.~W.,  {Teuben} P.,  2003, \apj, 582, 723

\bibitem[\protect\citeauthoryear{{Sakamoto}, {Okumura}, {Ishizuki} \&
  {Scoville}}{{Sakamoto} et~al.}{1999}]{1999ApJS..124..403S}
{Sakamoto} K.,  {Okumura} S.~K.,  {Ishizuki} S.,    {Scoville} N.~Z.,  1999,
  \apjs, 124, 403

\bibitem[\protect\citeauthoryear{{Sanders} \& {Huntley}}{{Sanders} \&
  {Huntley}}{1976}]{1976ApJ...209...53S}
{Sanders} R.~H.,  {Huntley} J.~M.,  1976, \apj, 209, 53

\bibitem[\protect\citeauthoryear{{Sandstrom} et~al.,}{{Sandstrom}
  et~al.}{2010}]{Sandstrom2010}
{Sandstrom} K.,  et~al., 2010, \aap, 518, L59+

\bibitem[\protect\citeauthoryear{{Schoenmakers}, {Franx} \& {de
  Zeeuw}}{{Schoenmakers} et~al.}{1997}]{Schoenmakers1997}
{Schoenmakers} R.~H.~M.,  {Franx} M.,    {de Zeeuw} P.~T.,  1997, \mnras, 292,
  349

\bibitem[\protect\citeauthoryear{{Schwarz}}{{Schwarz}}{1984}]{1984MNRAS.209...%
93S}
{Schwarz} M.~P.,  1984, \mnras, 209, 93

\bibitem[\protect\citeauthoryear{{Sersic}}{{Sersic}}{1958}]{sersic58}
{Sersic} J.~L.,  1958, The Observatory, 78, 123

\bibitem[\protect\citeauthoryear{{Shlosman}, {Frank} \& {Begelman}}{{Shlosman}
  et~al.}{1989}]{Shlosman1989}
{Shlosman} I.,  {Frank} J.,    {Begelman} M.~C.,  1989, \nat, 338, 45

\bibitem[\protect\citeauthoryear{{Storchi-Bergmann}, {Baldwin} \&
  {Wilson}}{{Storchi-Bergmann} et~al.}{1993}]{1993ApJ...410L..11S}
{Storchi-Bergmann} T.,  {Baldwin} J.~A.,    {Wilson} A.~S.,  1993, \apjl, 410,
  L11

\bibitem[\protect\citeauthoryear{{Storchi-Bergmann}, {Nemmen da Silva},
  {Eracleous}, {Halpern}, {Wilson}, {Filippenko}, {Ruiz}, {Smith} \&
  {Nagar}}{{Storchi-Bergmann} et~al.}{2003}]{StorchiBergmann2003}
{Storchi-Bergmann} T.,  {Nemmen da Silva} R.,  {Eracleous} M.,  {Halpern}
  J.~P.,  {Wilson} A.~S.,  {Filippenko} A.~V.,  {Ruiz} M.~T.,  {Smith} R.~C.,
   {Nagar} N.~M.,  2003, \apj, 598, 956

\bibitem[\protect\citeauthoryear{{Telesco} \& {Decher}}{{Telesco} \&
  {Decher}}{1988}]{1988ApJ...334..573T}
{Telesco} C.~M.,  {Decher} R.,  1988, \apj, 334, 573

\bibitem[\protect\citeauthoryear{{Tremaine} \& {Weinberg}}{{Tremaine} \&
  {Weinberg}}{1984}]{Tremaine1984}
{Tremaine} S.,  {Weinberg} M.~D.,  1984, \apjl, 282, L5

\bibitem[\protect\citeauthoryear{{Tully}}{{Tully}}{1988}]{1988ngc..book.....T}
{Tully} R.~B.,  1988

\bibitem[\protect\citeauthoryear{{van de Ven} \& {Chang}}{{van de Ven} \&
  {Chang}}{2009}]{vandeVenChang2009}
{van de Ven} G.,  {Chang} P.,  2009, \apj, 697, 619

\bibitem[\protect\citeauthoryear{{van de Ven} \& {Fathi}}{{van de Ven} \&
  {Fathi}}{2010}]{vandeVenFathi2010}
{van de Ven} G.,  {Fathi} K.,  2010, \apj, 723, 767

\bibitem[\protect\citeauthoryear{{van der Kruit} \& {Freeman}}{{van der Kruit}
  \& {Freeman}}{2011}]{2011ARA&A..49..301V}
{van der Kruit} P.~C.,  {Freeman} K.~C.,  2011, \araa, 49, 301

\bibitem[\protect\citeauthoryear{{Wada}}{{Wada}}{1994}]{Wada1994}
{Wada} K.,  1994, \pasj, 46, 165

\bibitem[\protect\citeauthoryear{{Wolstencroft} \& {Schempp}}{{Wolstencroft} \&
  {Schempp}}{1979}]{1979NZJS...22..325W}
{Wolstencroft} R.~D.,  {Schempp} W.~V.,  1979, New Zealand Journal of Science,
  22, 325

\end{thebibliography}

\label{lastpage}

\end{document}